\documentclass[a4paper]{article}
 
\usepackage{threeparttable,tabularx}
\usepackage{amsmath,amssymb}
\usepackage{booktabs}
\usepackage{multirow} 
\usepackage{algorithm}
\usepackage{graphicx}
\usepackage{algorithmicx}
\newcommand{\StepText}[2]{%
  \noindent
  \hangindent=3em
  \hangafter=1
  \textbf{#1} #2\par
}
\newcommand{\SubStep}[1]{%
  {\parindent=3.5em%
   \hangindent=5.5em%
   \hangafter=1%
   #1\par}
}

\newtheorem{theorem}{Theorem}

\newtheorem{corollary}{Corollary}
\newtheorem{lemma}{Lemma}
\newtheorem{assumption}{Assumption}
\newtheorem{proposition}{Proposition}

\usepackage[authoryear]{natbib}
\usepackage{hyperref} 
\usepackage{bbm}
\usepackage{pifont}
\usepackage[a4paper,
  top=2.5cm,
  bottom=2.5cm,
  left=2.8cm,
  right=2.8cm]{geometry}

\begin{document}

\title {Deep learning based doubly robust test for Granger causality\thanks{We thank Max H. Farrell for helpful comments.}} 

\author{
  Yongchang Hui$^{1}$,
  Chijin Liu$^{1}$,
  Xiaojun Song$^{2}$\thanks{Corresponding author} \\[6pt]
  {\small $^{1}$ Xi'an Jiaotong University, Xi'an, China} \\
  {\small $^{2}$ Peking University, Guanghua School of Management, Beijing, China}
}


\date{ }

\maketitle

\begin{abstract}

Granger causality is popular for analyzing time series data in many applications from natural science to social science including genomics, neuroscience, economics, and finance. Consequently, the Granger causality test has become one of the main concerns of the econometrician for decades. Taking advantage of the theoretical breakthroughs in deep learning in recent years, we propose a doubly robust Granger causality test (DRGCT). Our method offers several key advantages. The first and most direct benefit is for the users, DRGCT allows them to handle large lag orders while alleviating the curse of dimensionality that traditional nonlinear Granger causality tests usually face. Second, introducing a doubly robust test statistic for time series based on neural networks that achieves a parametric convergence rate not only suggests a new paradigm for nonparametric inference in econometrics, but also broadens the application scope of deep learning. Third, a multiplier bootstrap method, combined with the doubly robust approach, provides an efficient way to obtain critical values, effectively reducing computational time and avoiding redundant calculations. We prove that the test asymptotically controls the type I error, while achieving power approaches one, and validate the effectiveness of our test through numerical simulations. In real data analysis, we apply DRGCT to revisit the price-volume relationship problem in the stock markets of America, China, and Japan.

\ 


\end{abstract}

\section{Introduction}
Due to it's ability to process and analyze large complex datasets, the rapid development of deep learning has injected new momentum into economic and financial studies. The applications of deep learning range from classical demand estimation (\citeauthor{chernozhukov2019demand}, \citeyear{chernozhukov2019demand}) and solving dynamic economic models (\citeauthor{maliar2021deep}, \citeyear{maliar2021deep}) to  risk management and volatility prediction (\citeauthor{sadhwani2021deep}, \citeyear{sadhwani2021deep}; \citeauthor{liu2023deep}, \citeyear{liu2023deep}; \citeauthor{stok2024deep}, \citeyear{stok2024deep}),  asset pricing (\citeauthor{nagel2021machine}, \citeyear{nagel2021machine}; \citeauthor{gu2021autoencoder}, \citeyear{gu2021autoencoder}; \citeauthor{chen2024deep}, \citeyear{chen2024deep}) and portfolio allocation (\citeauthor{caner2022deep}, \citeyear{caner2022deep}). Deep learning has proven effective in tackling high-dimensional nonlinear modeling problems, which is also a key concern for econometricians. Pioneers have actively explored the value of deep learning in econometric theory (\citeauthor{farrell2021deep}, \citeyear{farrell2021deep}; \citeauthor{chernozhukov2022automatic}, \citeyear{chernozhukov2022automatic}; \citeauthor{hyun2025neural}, \citeyear{hyun2025neural}), especially the potential to address long-standing theoretical challenges.


In this study, we focus on the Granger causality introduced by \cite{granger1969investigating}, which is a mathematical framework with significant practical guidance to analyze dynamic relationships between time series. \cite{white2010granger} provided rigorous structural characterizations of Granger causality to avoid misuse of Granger’s pioneering concept. Predictability is a key feature in the Granger causality analysis and thus draws enormous attention from financial market participants and economic policy makers. Applications include the detection of financial risk (\citeauthor{hong2009granger}, \citeyear{hong2009granger}; \citeauthor{balboa2015granger}, \citeyear{balboa2015granger}; \citeauthor{corsi2018measuring}, \citeyear{corsi2018measuring}), as well as extensive research on the discovery of interrelationships between economic variables (\citeauthor{becker2013not}, \citeyear{becker2013not}; \citeauthor{puente2015granger}, \citeyear{puente2015granger}; \citeauthor{ramos2025granger}, \citeyear{ramos2025granger}).  \cite{white2014granger} presented a specific framework for economic policy analysis. Granger causality has also been widely applied in other various fields, from natural science (\citeauthor{shojaie2010discovering}, \citeyear{shojaie2010discovering}; \citeauthor{seth2015granger}, \citeyear{seth2015granger}; \citeauthor{abatis2024fear}, \citeyear{abatis2024fear}) to social science (\citeauthor{soysa1999boon}, \citeyear{soysa1999boon}; \citeauthor{seetanah2011assessing}, \citeyear{seetanah2011assessing}; \citeauthor{gunter2016forecasting}, \citeyear{gunter2016forecasting}). 

It is preferable that Granger causality methods are not constrained by a rigid model specification, since real-world data often exhibit complex nonlinear relationships (\citeauthor{tong1990non}, \citeyear{tong1990non}; \citeauthor{fan2003nonlinear}, \citeyear{fan2003nonlinear}; \citeauthor{granger2004time}, \citeyear{granger2004time}). Therefore, nonparametric approaches to the Granger causality test have received increasing attention. \cite{hidalgo2000nonparametric} introduced a nonparametric causality test in the frequency domain for weakly stationary linear processes, particularly for long-range-dependent observations. \cite{su2008nonparametric} proposed a nonparametric test for conditional independence based on the weighted Hellinger distance between conditional densities, which can be directly applied to testing Granger causality. \cite{nishiyama2011consistent} proposed a nonparametric test based on moment conditions, which allows nonlinear Granger causality. \cite{song2018measuring} developed model-free measures for Granger causality in mean and derived the asymptotic distribution of the associated nonparametric estimator, making it applicable for testing causality. 

Although nonparametric tests are highly flexible in detecting nonlinear Granger causality, the lag orders have to be chosen very small, which is a fundamental issue inherent in present methods. However, in most economic and financial application scenarios, the effects of a set of variables on others often take a while to work out. For example, the impact of economic policies or financial regulations may only become apparent after a considerable delay (\citeauthor{friedman1961lag}, \citeyear{friedman1961lag}; \citeauthor{jovanovski2011phenomenon}, \citeyear{jovanovski2011phenomenon}; \citeauthor{aruoba2024long}, \citeyear{aruoba2024long}). As the lag orders in Granger causality tests increase, traditional nonparametric methods  become ineffective even fail due to the curse of dimensionality. Considering the strong capacity to approximate complex nonlinear relationships, particularly exceptional performance in high-dimensional settings, neural network based learning for Granger causality is invented (e.g., \citeauthor{tank2021neural}, \citeyear{tank2021neural}; \citeauthor{nauta2019causal}, \citeyear{nauta2019causal}; \citeauthor{khanna2020economy}, \citeyear{khanna2020economy}). Learning for Granger causality opens up a new perspective for Granger causality study; however, these methods may not be suitable for econometric analysis, since it lacks support for statistical inference.


In recent years, theoretical advances have begun to emerge regarding the statistical inference of deep learning models. \cite{farrell2021deep} established the convergence rate of deep ReLU networks for bounded i.i.d. data. \cite{zhou2023testing} extended similar results to data that satisfy the $\beta$ mixing conditions, particularly for the mixture density network (MDN) introduced by \cite{bishop1994mixture}. \cite{brown2024statistical} further proposed the convergence rate of deep ReLU networks for unbounded dependent data. Our proposal relies on a multilayer perceptron (MLP) to estimate the conditional mean, and an MDN to effectively estimate conditional density distribution. Both components play an important role in the construction of the test statistic. Our test requires convergence rate guaranties for the MLP and the MDN based on semiparametric inference theories.


One should be aware that directly applying deep neural network estimators in the classic nonparametric test will lead to inevitable issues, such as excessive bias and the inability to control type I error. The root cause of these problems lies in the lack of established convergence rate guaranties for deep neural networks, making it impossible to ensure that the underlying stochastic process converges at an adequate rate. This, in turn, hinders the application of limit theorems to empirical processes in constructing the test statistic. To overcome this difficulty, we adopt a doubly robust structure, originally introduced by \cite{robins1994estimation}. The doubly robust framework has been extensively employed in causal inference and missing data problems, as it ensures consistency, provided that at least one of the involved models is correctly specified. See \cite{bang2005doubly} and \cite{vermeulen2015bias} for example. \cite{chernozhukov2018double} were among the first to introduce the doubly robust principle into the context of machine learning. More recently, \cite{zhou2023testing} developed a novel testing framework by incorporating the doubly robust principle into the study of Markov properties. Unlike in the missing data setting, when both models are well specified, the doubly robust approach in testing can further accelerate the convergence rate. 

Our proposal contributes to the literature in several meaningful ways. 
The most immediate advantage lies in practical applications. In real data analysis, selecting the appropriate lag order for Granger causality tests is often challenging, as researchers may have little guidance. The proposed DRGCT mitigates this issue by leveraging the strong performance of deep neural network estimators when handling multiple lag orders. As a result, one can choose a sufficiently long lag that encompasses all potential causal effects, thereby gaining a more comprehensive understanding of the underlying causal relationships.
In addition, we introduce deep learning techniques into the framework of the Granger causality test in a novel way. To apply deep neural network estimators within a classical statistical testing framework, we construct a doubly robust hypothesis structure to test Granger causality. We obtain a sharp upper bound on the approximation error for the MLP estimator under the $\beta$ mixing condition for both bounded and unbounded data based on \cite{farrell2021deep} and \cite{brown2024statistical}. By combining the convergence rate of the MDN estimator as in \cite{zhou2023testing}, we demonstrate that the proposed test successfully controls type I error and achieves a power approaching one. In particular, we establish theoretical validity for testing with unbounded data using deep learning–based estimators. In this sense, our work extends the applicability of hypothesis testing by incorporating deep learning tools. Furthermore, our test statistic achieves a parametric convergence rate, while its components are estimated semiparametrically. The difference in results between the doubly robust statistic and the single-estimator statistic reflects its effect on controlling the type I error. For local power analysis, tests based on empirical processes can achieve a parametric rate of $n^{-1/2}$, which is much faster than the rates obtained from smoothing-based nonparametric tests. It is worth noting that our proposed test does not rely on sample splitting or cross-fitting, as it effectively controls the type I error without these techniques. This offers the advantage of using the entire dataset for model training, thus improving the efficiency of estimation.
Finally, thanks to its doubly robust structure, the proposed test can employ a multiplier bootstrap method to obtain the critical value. Unlike the standard wild bootstrap, this procedure allows the critical value to be determined using the already estimated quantities, thereby avoiding redundant calculations. This feature is particularly important for deep learning–based methods, as it substantially reduces computational cost and running time.

The remainder of the article is organized as follows. Section \ref{hypo and test statistic} provides the motivation for the doubly robust test and introduces the corresponding hypothesis and the test statistic. Section \ref{section theory} presents the main theoretical results of the proposed test. Specifically, Section \ref{convergence rate of dnn} outlines the necessary lemmas that establish the convergence rates of the relevant deep neural network estimators, providing the theoretical foundation for our proposed test and reflecting the advantage of doubly robust construction when the convergence rates of individual estimators are insufficient. Section \ref{asymptotic null} establishes the asymptotic distribution of the test statistic under the null hypothesis, while Section \ref{alternatives} discusses the consistency of the test and its asymptotic local power under local alternatives. In Section \ref{bootstrap}, we present a bootstrap procedure for determining the critical value and implementing the testing procedure. Section \ref{simulations} reports the simulation results, including a comparison with the nonparametric test in \cite{nishiyama2011consistent}, and Section \ref{real data} illustrates the methodology using real data sets. Section \ref{conclusion} is the conclusion. All assumptions and technical proofs are provided in \hyperlink{myappendix}{Appendix}.

\section{Hypotheses and test statistic}
\label{hypo and test statistic}

Without loss of generality, consider a bivariate strictly stationary ergodic time series $\{\left(X_t, Y_t\right)\}$ defined on probability space $(\Omega, \mathcal{F}, {P})$, which satisfies the Markov property
\begin{equation}
\label{markov}
{P}\left(X_t \leq x, Y_t \leq y | \mathcal{F}_{t-1}\right)=\mathrm{P}\left(X_t \leq x, Y_t \leq y | \boldsymbol{W}_{t-1}\right),\quad \text { a.s. } \quad \forall(x, y) \in \mathbb{R}^{2},
\end{equation}
where $\mathcal{F}_{t-1} = \sigma \left(\{ X_{s},Y_{s}\}_{s=-\infty}^{t-1}\right)$ with $\sigma(\cdot)$ the smallest sigma algebra, and $\boldsymbol{W}_{t-1}=\left\{\left(\boldsymbol{X}_{t-1}^{\top}, \boldsymbol{Y}_{t-1}^{\top}\right)^{\top}\right\}$,
in which $\boldsymbol{X}_{t-1} = (X_{t-1},\cdots,X_{t-p})^\top$, $\boldsymbol{Y}_{t-1} = (Y_{t-1},\cdots,Y_{t-q})^\top$ with two integers $p,\;q\in (0,\infty)$. That is, the only relevant information to forecast $\left(X_t, Y_t\right)$ consists of the first $p$ lags of $X_{t}$ and the first $q$ lags of $Y_{t}$. Then, the conditional mean of $Y_t$ given $\mathcal{F}_{t-1}$ can be reduced to $\mathbb{E}(Y_t|\boldsymbol{W}_{t-1})$.

According to \cite{granger1969investigating}, if $X_t$ does not Granger cause $Y_t$ in mean, then
$\mathbb{E}(Y_t|\boldsymbol{W}_{t-1}) = \mathbb{E}(Y_t|\boldsymbol{Y}_{t-1})$.
Denote $\mathbb{E}(Y_t|\boldsymbol{Y}_{t-1}) = m(\boldsymbol{Y}_{t-1})$. Thus, the null hypothesis that $X_t$ does not Granger cause $Y_t$ can be reformulated as a conditional moment restriction, which has the advantage of conditioning only on $\boldsymbol{W}_{t-1}$. Therefore, our goal now is to test
\begin{equation}
\label{condi}
\mathrm{H_0}: \mathbb{E}\left[Y_t - m(\boldsymbol{Y}_{t-1})| \boldsymbol{W}_{t-1}\right]=0, \quad \mathrm{a.s.} 
\end{equation} 
The alternative hypothesis $\mathrm{H_1}$ is the negation of $\mathrm{H_0}$ in \eqref{condi}.
According to \cite{stinchcombe1998consistent}, it is equivalent to test
$$
\mathrm{H_0}: \mathbb{E}\left[\varphi\left(\boldsymbol{W}_{t-1}, w\right)\left(Y_t - m(\boldsymbol{Y}_{t-1})\right)\right] = 0,\ \  \forall {w}\in \mathcal{W},\quad \mathrm{a.s.}
$$
where $\varphi$ is a generically comprehensively revealing (GCR) or comprehensively revealing (CR) function, and $\mathcal{W} \subseteq \mathbb{R}^{p+q}$ is a properly chosen set accordingly. Candidates for GCR functions include $\varphi\left(\boldsymbol{W}_{t-1}, w\right)=\exp \left(i {w}^{\top} \boldsymbol{W}_{t-1}\right)$ and $\varphi\left(\boldsymbol{W}_{t-1}, w\right)=\sin \left(w^{\top} \boldsymbol{W}_{t-1}\right)$, while examples of CR functions include $\varphi\left(\boldsymbol{W}_{t-1}, w\right)=\mathbbm{1}\left(\boldsymbol{W}_{t-1} \leq w\right)$ and $\varphi\left(\boldsymbol{W}_{t-1}, w\right)=\mathbbm{1}\left(w^{\top} \boldsymbol{W}_{t-1} \leq \alpha\right)$. It is worth noting that when $\varphi$ is chosen from GCR functions, deviations from the null hypothesis can be detected for any choice of $w \in \mathcal{W}$, where $\mathcal{W}$ can be any small compact set with nonempty interior. In contrast, CR functions may require $\mathcal{W}$ to be the entire Euclidean space to ensure consistency of the test.  See \cite{stinchcombe1998consistent}, and \cite{su2012conditional} for further discussion. Hence, in the following, we choose the GCR function $\varphi\left(\boldsymbol{W}_{t-1}, w\right)=\exp \left({i} w^{\top} \boldsymbol{W}_{t-1}\right)$ for our test and assume that $\mathcal{W}$ is a compact set in $\mathbb{R}^{p+q}$.
This indicates that the null hypothesis $\mathrm{H}_0$ in \eqref{condi} is equally modified to
\begin{equation}
\label{uncondi}
\mathrm{H_0}: \mathbb{E}\left[\left(Y_t - m(\boldsymbol{Y}_{t-1})\right)e^{i{w}^\top\boldsymbol{W}_{t-1}}\right] = 0,\ \  \forall {w}\in \mathcal{W}, \quad \mathrm{a.s.}
\end{equation}

Given a sample of time series $\left\{ (X_t, Y_t) \right\}_{t=1}^{n}$, suppose $p\leq q$ hereinafter. Firstly, consider the infeasible empirical process
\begin{equation}
\label{Sn0}
\begin{aligned}
    S_n^0(\mu, \nu) &= \frac{1}{\sqrt{n-q}} \mathop{\sum}\limits_{t=q+1}^{n}\Big(Y_t-{m}(\boldsymbol{Y}_{t-1})\Big)e^{i{w}^\top\boldsymbol{W}_{t-1}}\\
    &=\frac{1}{\sqrt{n-q}} \mathop{\sum}\limits_{t=q+1}^{n}\Big(Y_t-{m}(\boldsymbol{Y}_{t-1})\Big)e^{i\mu^\top\boldsymbol{Y}_{t-1}}e^{i\nu^\top\boldsymbol{X}_{t-1}}.
\end{aligned}
\end{equation}
To ensure clarity, let $\mathcal{W} = \mathcal{W}_1 \times \mathcal{W}_2$, and $\mu \in \mathcal{W}_1$, $\nu \in \mathcal{W}_2$.
To construct the test, we need to substitute $m(\boldsymbol{Y}_{t-1})$ with its estimator. This suggests the empirical process
\begin{equation}
\label{Sn0hat}
\widehat{S}_n^0(\mu, \nu) =\frac{1}{\sqrt{n-q}} \mathop{\sum}\limits_{t=q+1}^{n}\Big(Y_t-\widehat{m}(\boldsymbol{Y}_{t-1})\Big)e^{i\mu^\top\boldsymbol{Y}_{t-1}}e^{i\nu^\top\boldsymbol{X}_{t-1}}.
\end{equation}
We depart from the settings in the classical nonlinear Granger causality test by allowing for large lag orders $p$ and $q$. Traditional nonparametric methods for estimating $m(\boldsymbol{Y}_{t-1})$ may fail due to "the curse of dimensionality". Therefore, we apply deep learning techniques, which are particularly well suited to estimate ${m}(\boldsymbol{Y}_{t-1})$ in this situation. However, from \cite{chernozhukov2018double}, A naive application of a deep neural network to estimate ${m}(\boldsymbol{Y}_{t-1})$ will cause a substantial bias in the test statistic based on \eqref{Sn0hat} under the null hypothesis.

To address this issue, we consider to construct a doubly robust test. Thus, we further manipulate the null hypothesis \eqref{uncondi} into
\begin{equation}
\label{drh}
\mathrm{H_0}: \mathbb{E} \left[ \Big(Y_t-{m}(\boldsymbol{Y}_{t-1})\Big)e^{i\mu^\top\boldsymbol{Y}_{t-1}}\left(e^{i\nu^\top\boldsymbol{X}_{t-1}}-{\phi}(\nu| \boldsymbol{Y}_{t-1})\right) \right] = 0 , \ \  \forall {(\mu,\nu)}\in \mathcal{W}, \quad \mathrm{a.s.}
\end{equation}
where ${\phi}(\nu| \boldsymbol{Y}_{t-1}) = \mathbb{E}\left[e^{i\nu^\top\boldsymbol{X}_{t-1}}| \boldsymbol{Y}_{t-1}\right]$. The verification of the equivalence between \eqref{uncondi} and \eqref{drh} is provided in Proposition \ref{hyp=} of Appendix \ref{main proofs}.

According to \eqref{drh}, define a new infeasible empirical process,
\begin{equation}
\label{infedr}
    S_n({\mu},{\nu}) = \frac{1}{\sqrt{n-q}} \mathop{\sum}\limits_{t=q+1}^{n}\Big(Y_t-{m}(\boldsymbol{Y}_{t-1})\Big)e^{i\mu^\top\boldsymbol{Y}_{t-1}}\left(e^{i\nu^\top\boldsymbol{X}_{t-1}}-{\phi}(\nu| \boldsymbol{Y}_{t-1})\right).
\end{equation}
Henceforth, let $\widehat{m}(\boldsymbol{Y}_{t-1})$ denote the estimator of $m(\boldsymbol{Y}_{t-1})$ obtained using the MLP. Following \cite{zhou2023testing}, we estimate ${\phi}(\nu|\boldsymbol{Y}_{t-1})$ through deep conditional generative learning and denote the estimator by $\widehat{\phi}(\nu| \boldsymbol{Y}_{t-1})$.
Then, our test statistic is based on the following feasible empirical process
\begin{equation}
\label{fedr}
    \widehat{S}_n({\mu},{\nu}) = \frac{1}{\sqrt{n-q}} \mathop{\sum}\limits_{t=q+1}^{n}\left(Y_t-\widehat{m}(\boldsymbol{Y}_{t-1})\right)e^{i\mu^\top\boldsymbol{Y}_{t-1}}\left(e^{i\nu^\top\boldsymbol{X}_{t-1}}-\widehat{\phi}(\nu| \boldsymbol{Y}_{t-1})\right).
\end{equation}
A key advantage is that ${S}_n({\mu},{\nu})-\widehat{S}_n({\mu},{\nu})$ decays to zero much faster than ${S}^0_n({\mu},{\nu})-\widehat{S}^0_n({\mu},{\nu})$ under the null hypothesis.

Before introducing the final test statistic, we outline the procedure for estimating $m(\boldsymbol{Y}_{t-1})$ and $\phi(\nu| \boldsymbol{Y}_{t-1})$ by deep learning techniques based on the MLP and MDN frameworks, respectively. 

The reorganized samples $\{Y_t, \boldsymbol{Y}_{t-1}\}_{t=q+1}^{n}$ are used to train the MLP and obtain conditional expectation estimators $\left\{\widehat{m}(\mathbf{Y}_{t-1})\right\}_{t=q+1}^n$, where $\boldsymbol{Y}_{t-1}$ and $Y_t$ are the input covariates and the response variable, respectively. For estimating $\phi(\nu| \boldsymbol{Y}_{t-1})$, we begin with inputting $\{\boldsymbol{X}_{t-1}, \boldsymbol{Y}_{t-1}\}_{t=q+1}^n$ to train the estimated conditional density function $\widehat{f}_{\boldsymbol{X}_{t-1}| \boldsymbol{Y}_{t-1}}(x|y)$ by the MDN with $G$ mixture components, where $\boldsymbol{Y}_{t-1}$ and $\boldsymbol{X}_{t-1}$ are the input and response, respectively. For each $t\in \{q+1, q+2, \cdots,n\}$, given $\boldsymbol{Y}_{t-1}$, we randomly generate $M$ samples $\left\{\boldsymbol{X}_{j}^*\right\}_{j=1}^M$ by $\widehat{f}_{\boldsymbol{X}_{t-1}| \boldsymbol{Y}_{t-1}}(x|y)$. Let $\widehat{\phi}\left(\nu| \boldsymbol{Y}_{t-1}\right)=\frac{1}{M} \sum_{j=1}^M \exp \left(i \nu^\top \boldsymbol{X}_{j}^*\right)$. Then, we obtain $\left\{\widehat{\phi}\left(\nu| \boldsymbol{Y}_{t-1}\right)\right\}_{t=q+1}^n$ for some $\nu \in \mathcal{W}_2$.
For details on the MDN framework, see \cite{bishop1994mixture} or \cite{zhou2023testing}. The estimation procedures are summarized as \textbf{Step 1} and \textbf{Step 2} in Algorithm~\ref{algo}.

To construct the test statistic, we consider different combinations of $(\mu, \nu)$ for $\widehat{S}_n(\mu, \nu)$ in \eqref{fedr}. Hence, we randomly sample $L$ i.i.d. pairs $\left\{\left(\mu_l, \nu_l\right)\right\}_{l=1}^L$ from a multivariate uniform distribution over a specified compact interval. 
Then, we obtain the discretized version of $\widehat{S}_n (\mu,\nu)$, that is
\begin{equation}
\label{Sn hat mu nu}
    \widehat{S}_n({\mu_l},{\nu_l}) = \frac{1}{\sqrt{n-q}} \mathop{\sum}\limits_{t=q+1}^n\left(Y_t-\widehat{m}(\boldsymbol{Y}_{t-1})\right)e^{i\mu_l^\top\boldsymbol{Y}_{t-1}}\left(e^{i\nu_l^\top\boldsymbol{X}_{t-1}}-\widehat{\phi}(\nu_l| \boldsymbol{Y}_{t-1})\right).
\end{equation}
Since $\widehat{S}_n({\mu_l},{\nu_l})$ is complex valued, we decompose it into two parts as real and imaginary parts, denoted $\widehat{S}_R({\mu_l},{\nu_l})$ and $\widehat{S}_I({\mu_l},{\nu_l})$, respectively. Finally, we propose the \textbf{Kolmogorov–Smirnov test statistic} as
\begin{equation}
\label{KSn}
\mathrm{KS}_n=\max _{l \in\{1, \ldots, L\}} \  \max \left\{\lvert \widehat{S}_R\left(\mu_l, \nu_l\right)\rvert , \lvert \widehat{S}_I\left(\mu_l, \nu_l\right)\rvert \right\} .
\end{equation}
The testing procedures are summarized as \textbf{Step 3},  \textbf{Step 4} and \textbf{Step 5} in Algorithm~\ref{algo}, with the bootstrap procedure explained in Section~\ref{bootstrap}.

\begin{algorithm}[htbp]
    \caption{Estimation and testing procedures of the doubly robust test for Granger causality}
    \label{algo}
    \StepText{Input:}{Data $\{X_t, Y_t\}_{t=1}^{n}$, the corresponding numbers of lags $p$ and $q$, the number of mixture components of the MDN $G$, the number of pairs $L$, the number of samples from the generators $M$, the number of bootstrap replications $B$, and the significance level $\alpha$.
    }

    \vspace{.5\baselineskip}
     
    \StepText{Step 1:}{
    Derive the conditional expectation estimators.}
      
    \SubStep{
    (1a) Reorganize part of the input data into $\{Y_t, \boldsymbol{Y}_{t-1}\}_{t=q+1}^{n}$.
    }
      
    \SubStep{
    (1b) Use the lagged observations $\boldsymbol{Y}_{t-1}$ as input covariates and $Y_t$ as the response variable to train the MLP model, and obtain estimated conditional expectations $\left\{\widehat{m}(\boldsymbol{Y}_{t-1})\right\}_{t=q+1}^n$.
    }
    
    \vspace{.5\baselineskip}

    \StepText{Step 2:}{
    Derive the conditional characteristic function estimators.
    }

    \SubStep{
    (2a) Reorganize the input data into $\{\boldsymbol{X}_{t-1}, \boldsymbol{Y}_{t-1}\}_{t=q+1}^n$.
    }

    \SubStep{
    (2b) Use $\boldsymbol{Y}_{t-1}$ as input covariates and $\boldsymbol{X}_{t-1}$ as the response variable to train the MDN model with $G$ mixture components and obtain the estimated conditional density $\widehat{f}_{\boldsymbol{X}_{t-1}| \boldsymbol{Y}_{t-1}}(x|y)$.
    }

    \SubStep{
    (2c) Given $\boldsymbol{Y}_{t-1}$, randomly generate $M$ samples $\left\{\boldsymbol{X}_{j}^*\right\}_{j=1}^M$ through $\widehat{f}_{\boldsymbol{X}_{t-1}| \boldsymbol{Y}_{t-1}}(x|y)$.
    }

    \SubStep{
    (2d) Randomly sample $L$ i.i.d. pairs $\left\{\left(\mu_l, \nu_l\right)\right\}_{l=1}^L$ from a multivariate uniform distribution over a  compact interval.
    }

    \SubStep{
    (2e) Calculate the conditional characteristic function estimators $\widehat{\phi}\left(\nu_l| \boldsymbol{Y}_{t-1}\right)=\frac{1}{M} \sum_{j=1}^M \exp \left(i \nu_l^\top \boldsymbol{X}_{j}^*\right)$ for $l=1, \cdots, L$ and $t=q+1, \cdots, n$.
    }

    \vspace{.5\baselineskip}

    \StepText{Step 3:}{
    Construct the Kolmogorov–Smirnov test statistic.
    }

    \SubStep{
    (3a) Compute $\widehat{S}_n(\mu_l,\nu_l)$ according to \eqref{Sn hat mu nu}.
    }

    \SubStep{
    (3b) Construct the Kolmogorov–Smirnov test statistic $\mathrm{KS}_n$ according to \eqref{KSn}.
    }

    \vspace{.5\baselineskip}

    \StepText{Step 4:}{
    Derive the critical value through a bootstrap procedure.
    }

    \SubStep{
    (4a) Generate i.i.d samples $\left\{\xi_t\right\}_{t=1}^n$ from a distribution of zero mean and unit variance, independent of $\left\{(X_t,Y_t)\right\}_{t=1}^n$. For each $l \in \{ 1,\cdots, L\}$, compute $\widehat{S}_n^*\left(\mu_l, \nu_l\right)$ according to \eqref{Snstar} and obtain $\mathrm{KS}_n^*$.
    }
    
    \SubStep{
    (4b) Repeat {(4a)} for $B$ times to obtain $\left\{\mathrm{KS}_{n, b}^*\right\}_{b=1}^B$, and compute the bootstrapped $p$-value $p_n^*$ by the equation $p_n^*=B^{-1} \sum_{b=1}^B \mathbbm{1}\left(\mathrm{KS}_{n, b}^* \geq \mathrm{KS}_n\right)$.
    }

    \vspace{.5\baselineskip}
    
    \StepText{Step 5:}{
    Reject $\mathrm{H}_0$ if $p_n^*<\alpha$.
    }

\end{algorithm} 

\section{Theory}
\label{section theory}

In this section, we first discuss the convergence rates of the deep neural network based estimators $\widehat{m}(\boldsymbol{Y}_{t-1})$ and $\widehat{\phi}(\nu|\boldsymbol{Y}_{t-1})$, which serve as the theoretical basis for the proposed test. Then, we present the asymptotic properties of the empirical processes (\ref{fedr}) and the test statistic $\mathrm{KS}_n$ (\ref{KSn}) under the null and alternatives in Section \ref{asymptotic null} and \ref{alternatives}, respectively. Finally, we obtain the critical value of the test through a multiplier bootstrap procedure and verify its validity.

\subsection{Convergence rates of deep neural networks}
\label{convergence rate of dnn}

Before establishing the asymptotic properties of the empirical process and the corresponding test statistic, we first derive the sharp upper bounds for the deep neural network estimators $\widehat{m}(\boldsymbol{Y}_{t-1})$ and $\widehat{\phi}(\nu|\boldsymbol{Y}_{t-1})$. Henceforth, the notations are in \hyperref[appendixbegin]{Appendix}, the assumptions are stated in Appendix \ref{assumptions}, and the proofs of lemmas are presented in Appendix \ref{tech lemmas}.

The following two lemmas provide the convergence rates of the MLP estimator $\widehat{m}(\boldsymbol{Y}_{t-1})$ for bounded and unbounded $Y_t$, corresponding to Assumption \ref{boundv} and \ref{unboundv}, respectively.
\begin{lemma}
\label{mlpL2}
Suppose Assumptions \ref{betamixing}, \ref{boundv} and \ref{mlpf} hold. Then, there exists a MLP estimator $\widehat{m}(\boldsymbol{Y}_{t-1})$ for $m(\boldsymbol{Y}_{t-1})$, with width $H_n \asymp n^{\frac{q}{2(\beta_0+q)}} \log ^2 (n)$ and depth $L_n \asymp \log (n)$, such that
$$
\mathbb{E} \left( \widehat{m}(\boldsymbol{Y}_{t-1})-m(\boldsymbol{Y}_{t-1})\right)^2 \leq C \cdot\left\{n^{-\frac{\beta_0}{\beta_0+q}} \log ^9 (n)+\frac{\log (n) \log \log (n)}{n}\right\},
$$
and
$$
\mathbb{E}_n \left( \widehat{m}(\boldsymbol{Y}_{t-1})-m(\boldsymbol{Y}_{t-1})\right)^2 \leq C \cdot\left\{n^{-\frac{\beta_0}{\beta_0+q}} \log ^9 (n)+\frac{\log (n) \log \log (n)}{n}\right\},
$$
for a constant $C>0$, with probability at least $1-4\exp \left(-n^{\frac{q}{\beta_0+q}} \log ^8 (n)\right)-O(n^{-1})$, where $\mathbb{E}_n$ denotes the empirical expectation.
\end{lemma}

Next, we remove the bounded restriction in Assumption \ref{boundv}, based on
the theories in \cite{brown2024statistical}. Lemma~\ref{mlpL2un} provides theoretical justification for the doubly robust test in the presence of unbounded dependent data. 

\begin{lemma}
\label{mlpL2un} Suppose Assumptions \ref{betamixing},  \ref{unboundv} and \ref{mlpf} hold. Suppose the MLP is specified with width
$H_n \asymp n^{\left(\frac{q}{\beta_0+q}\right)\left(1 / 2-\kappa_B\right)} \log ^2(n)$ and depth $L_n \asymp \log (n)$. Let $
K_n=n^{-\left(\frac{\beta_0}{\beta_0+q}\right)\left(1 / 2-\kappa_B\right)} \log ^6(n)$.
Then, for the MLP estimator $\widehat{m}(\boldsymbol{Y}_{t-1})$, 
there exist constants $C, C_1, C_2>0$ independent of $n$, for all $n$ sufficiently large, such that
$$
\begin{aligned}
& P\left(\mathbb{E}\left|\widehat{m}(\boldsymbol{Y}_{t-1})-m(\boldsymbol{Y}_{t-1})\right|^2 \leq C^2 K_n^2\right) \geq 1-e^{-n\left(\frac{\beta_0}{\beta_0+q}\right)^{\left(1 / 2-\kappa_B\right)}}-\frac{2 C_1 n^{1-C_2 \log (n)}}{\log (n)}-o(1/n), \\
& P\left(\mathbb{E}_n\left |\widehat{m}(\boldsymbol{Y}_{t-1})-m(\boldsymbol{Y}_{t-1})\right |^2 \leq C^2 K_n^2\right) \geq 1-4 e^{-n\left(\frac{\beta_0}{\beta_0+q}\right)^{\left(1 / 2-\kappa_B\right)}}-\frac{12 C_1 n^{1-C_2 \log (n)}}{\log (n)}-o(1/n).
\end{aligned}
$$

\end{lemma}

In fact, some mild conditions are not stated in detail in Lemma~\ref{mlpL2un}, including one on the MLP architecture and another on the training accuracy. For more details, see (2.2) and (3.2) in \cite{brown2024statistical}.

Next, we establish the error bound of the mixture density network estimator.
Let $f_{\boldsymbol{X}_{t-1} | \boldsymbol{Y}_{t-1}}(x|y)$ denote the true conditional density function of $\boldsymbol{X}_{t-1}$ given $\boldsymbol{Y}_{t-1}$.
\begin{lemma}
\label{mdnL2}
    Suppose Assumptions \ref{betamixing} and \ref{smooth} hold.  Then, there exists a certain MDN function class satisfying Assumption \ref{mdncondition}, such that the resulting MDN estimator $\widehat{f}_{\boldsymbol{X}_{t-1} | \boldsymbol{Y}_{t-1}}(x|y)$ satisfies that
$$
\begin{aligned}
\mathbb{E}\left( \widehat{f}_{\boldsymbol{X}_{t-1} | \boldsymbol{Y}_{t-1}}(x|y)-f_{\boldsymbol{X}_{t-1} | \boldsymbol{Y}_{t-1}}(x|y)\right)^2 & ={\int_{x, y}\lvert \widehat{f}_{\boldsymbol{X}_{t-1} | \boldsymbol{Y}_{t-1}}(x | y)-f_{\boldsymbol{X}_{t-1} | \boldsymbol{Y}_{t-1}}(x | y)\rvert ^2 d x d y} \\
& \leq C q\left\{G^{-\omega_1}+G^{\frac{\gamma_0+q}{2 \gamma_0}+4 \omega_2} n^{-\frac{\gamma_0}{2 \gamma_0+q}} \log ^3(n G)\right\},
\end{aligned}
$$
for some constants $C, \omega_1, \omega_2>0$, with probability at least $1-O\left(n^{-1}\right)$.
\end{lemma}

Based on Lemma~\ref{mdnL2}, the following lemma provides the convergence rate of the estimator $\widehat{\phi}(\cdot|\cdot)$.
\begin{lemma}
\label{phiL2}
    Suppose Assumptions \ref{betamixing} and \ref{convergencerate} (ii)-(iv) hold. Then, there exists a constant $\kappa>1 / 2$, such that
$$
\begin{gathered}
\max _{1 \leq l \leq L} \int_y\left\lvert\widehat{\phi}\left(\nu_l | y\right)-\phi\left(\nu_l | y\right)\right\rvert^2 \mathbb{F}(\mathrm{d} y)=O_p\left(n^{-\kappa}\right) \\
\end{gathered}
$$
where $\mathbb{F}$ denotes the cumulative distribution function of $\boldsymbol{Y}_{t-1}$, and ${\phi}(\nu|y) = \mathbb{E}\left(e^{i\nu^\top \boldsymbol{X}_{t-1}}|\boldsymbol{Y}_{t-1}=\boldsymbol{y}\right)$.
\end{lemma}

The lemmas \ref{mlpL2}-\ref{phiL2} indicate that the convergence rates of deep neural network estimators fall short of the parametric rate. Therefore, employing a doubly robust construction in the test ensures a sufficiently fast convergence rate for establishing the limiting theorems in Section \ref{asymptotic null}.

\subsection{Asymptotic null distributions}
\label{asymptotic null}

In this section, we establish the asymptotic null distribution of the test statistic $\mathrm{KS}_n$ in \eqref{KSn}. The proofs of theorems are presented in Appendix \ref{main proofs}.
First, we show that $\widehat{S}_n(\mu, \nu)$ defined in \eqref{fedr} converges weakly to a Gaussian process under the null hypothesis.

\begin{theorem}
\label{theoremnull}
Suppose Assumptions \ref{betamixing}, \ref{convergencerate} and \ref{distribution} hold. Then, under the null hypothesis in \eqref{drh},
$$
\widehat{S}_n(\cdot, \cdot) \rightsquigarrow S_{\infty}(\cdot, \cdot),
$$
where $S_{\infty}(\cdot, \cdot)$ is a zero mean Gaussian process with covariance kernel $\mathbb{E}\left[S_{\infty}(\mu, \nu)S_{\infty}(\mu', \nu')\right]$ and "$\rightsquigarrow$" denotes the weak convergence.
\end{theorem}

It worth noting that subtracting the term ${\phi}(\nu| \boldsymbol{Y}_{t-1})$ from $e^{i\nu^\top\boldsymbol{X}_{t-1}}$ in \eqref{infedr} is crucial for fixing the cost on convergence rate incurred by replacing ${m}(\boldsymbol{Y}_{t-1})$ with the MLP estimator $\widehat{m}(\boldsymbol{Y}_{t-1})$ in \eqref{Sn0}, known as the estimation error. The construction of ${S}_n(\mu,\nu)$ ensures that ${S}_n(\mu,\nu) - \widehat{S}_n(\mu,\nu)$ depends on the product of the convergence rates of the two estimators $\widehat{m}$ and $\widehat{\phi}$ under the null hypothesis. This double robustness property allows us to employ deep learning based estimators, as shown in Step 1 of the proof of Proposition \ref{mainpro}.   
On the contrary, the bias between $\widehat{S}_n^0(\mu, \nu)$ and ${S}_n^0(\mu, \nu)$ only depends on $\widehat{m}$. This leads to higher requirements on the convergence rate of $\widehat{m}$, which obstructs us from using the deep neural network estimator. In fact, $\widehat{S}_n^0(\mu, \nu)$ in \eqref{Sn0hat} does not converge in distribution to a Gaussian process under the null hypothesis.

The asymptotic null distribution of the test statistic $\mathrm{KS}_n$ in \eqref{KSn} is provided in the following corollary.

\begin{corollary}
\label{conull}
Suppose Assumptions \ref{betamixing}, \ref{convergencerate} and \ref{distribution} hold. Then, under the null hypothesis in \eqref{drh},
$$
\mathrm{KS}_n \rightsquigarrow \mathrm{KS}_{\infty}=\sup _{(\mu,\nu) \in \mathcal{W}} \max\left(| S_{R}(\mu, \nu)| , | S_{I}(\mu, \nu)| \right),
$$
where $S_{R}(\mu, \nu)$ and $S_{I}(\mu, \nu)$ denote the real and imaginary parts of $S_{\infty}(\mu, \nu)$, respectively.
\end{corollary}

Although we have provided the limiting distribution of the test statistic under the null hypothesis, it is still difficult to obtain the critical value of the test given the significance level $\alpha$. Therefore, we develop a bootstrap procedure for our test. See Section \ref{bootstrap} for details.

\subsection{Consistency and asymptotic local power}
\label{alternatives}

To investigate the consistency and asymptotic local power of our test, we further derive the asymptotic distribution of $\widehat{S}_n(\mu, \nu)$ in \eqref{fedr} under the alternative and the sequence of local alternatives defined in \eqref{h1n}.

\begin{theorem}
\label{theoremal}
Suppose Assumptions \ref{betamixing}, \ref{convergencerate} and \ref{distribution} hold. Then, under the alternative hypothesis, for each $(\mu, \nu) \in \mathcal{W}$,
$$
n^{-1/2} \widehat{S}_n(\mu, \nu) \stackrel{P}{\rightarrow} \mathbb{E}\left[\left(Y_t-m(\boldsymbol{Y}_{t-1})\right)e^{iw^\top \boldsymbol{W}_{t-1}}\right].
$$
\end{theorem}
Under the alternative, since $\mathbb{E}\left[\left(Y_t-m(\boldsymbol{Y}_{t-1})\right)e^{iw^\top \boldsymbol{W}_{t-1}}\right] \neq 0$, our statistic $\mathrm{KS}_n$ diverges to infinity and achieves asymptotic power of one.

To analyze the asymptotic local power property of the test, we introduce a sequence of local alternatives, that is
\begin{equation}
\label{h1n}
\mathrm{H}_{1n}: \mathbb{E}\left[Y_t - m(\boldsymbol{Y}_{t-1}) | \boldsymbol{W}_{t-1} \right]= n^{-1/2}\Delta(\boldsymbol{W}_{t-1}), \quad \mathrm{a.s.}
\end{equation}
where $\Delta$ is a non-constant measurable function, satisfying $Q(\mu,\nu)=\mathbb{E}(\Delta(\boldsymbol{W}_{t-1})e^{iw^\top\boldsymbol{W}_{t-1}})\neq 0$ a.s. for $w\in \mathcal{W}$. Recall that $w=(\mu,\nu)$. The form of local alternatives in \eqref{h1n} is commonly used when studying the asymptotic local power of a test based on empirical processes.

In \eqref{h1n}, $n^{-1/2}\Delta(\boldsymbol{W}_{t-1})$ characterizes the deviation of $Y_t$ from the conditional mean $m(\boldsymbol{Y}_{t-1})$ given $\boldsymbol{W}_{t-1}$. Specifically, $\Delta(\boldsymbol{W}_{t-1})$ determines the direction of the departure, while $n^{-1/2}$ represents the rate at which the difference between $Y_t$ and $m(\boldsymbol{Y}_{t-1})$. 

To derive the local power results, an additional condition is required, that is Assumption \ref{local_alternative}, stated in Appendix \ref{assumptions}.
The following theorem shows the asymptotic behavior of $\widehat{S}_n(\mu, \nu)$ under the sequence of local alternatives in \eqref{h1n}.

\begin{theorem}
\label{theoremlo}
Suppose Assumptions \ref{betamixing}, \ref{convergencerate}, \ref{distribution} and \ref{local_alternative} hold. Then, under the sequence of local alternatives in \eqref{h1n},
$$
\widehat{S}_n(\cdot, \cdot) \rightsquigarrow S_{\infty}(\cdot, \cdot)+Q(\cdot, \cdot),
$$
where $S_{\infty}(\cdot, \cdot)$ is the Gaussian process defined in Theorem \ref{theoremnull} and $Q(\cdot, \cdot)$ is the shift function.
\end{theorem}

Since the shift function $Q(\mu,\nu)  \neq 0$ a.s. , the test statistic $\mathrm{KS}_n$, based on $\widehat{S}_n(\mu,\nu)$, will have non-trivial local power against $\mathrm{H}_{1 n}$. Note that $\mathrm{H}_{1 n}$ converges to $\mathrm{H}_0$ at a parametric rate $n^{-1 / 2}$, the fastest rate known to test the Granger causality in mean.
For local power analysis, a key advantage of the empirical processes-based test is to achieve the parametric rate $n^{-1/2}$, which is significantly faster than the rates typically attained by the smoothing-based nonparametric test.

The limiting distribution of $\mathrm{KS}_n$ under the sequence of local alternatives in \eqref{h1n} is stated in the following corollary.

\begin{corollary}
\label{colo}
Suppose Assumptions \ref{betamixing}, \ref{convergencerate}, \ref{distribution} and \ref{local_alternative} hold. Then, under the sequence of local alternatives in \eqref{h1n},
$$
\begin{gathered}
\mathrm{KS}_n \rightsquigarrow \sup _{(\mu, \nu) \in \mathcal{W}} \max\left(| S_{R}(\mu, \nu) + Q_{R}(\mu, \nu)| , | S_{I}(\mu, \nu)+Q_{I}(\mu, \nu)| \right) , 
\end{gathered}
$$
where $Q_{R}(\mu, \nu)$ and $Q_{I}(\mu, \nu)$ denote the real and imaginary parts of $Q(\mu, \nu)$, respectively.
\end{corollary}

\subsection{Bootstrap}
\label{bootstrap}
Due to the complicated covariance kernel of the limiting process $S_{\infty}(\mu,\nu)$ in Theorem \ref{theoremnull}, the asymptotic null distribution of $\mathrm{KS}_n$ depends on the underlying data-generating process. This prevents us from directly obtaining the critical values of the test. To address this issue, we propose a bootstrap procedure inspired by the multiplier bootstrap method suggested by \cite{delgado2001significance}. This procedure fully leverages the asymptotic theory established in Theorem \ref{theoremnull} and is computationally efficient, as it avoids recalculating estimates in each bootstrap replication.

Now, consider the multiplier bootstrap empirical process
\begin{equation}
\label{Snstar}
\widehat{S}_n^*(\mu,\nu) = \frac{1}{\sqrt{n-q}} \mathop{\sum}\limits_{t=q+1}^n \xi_t \left(Y_t-\widehat{m}(\boldsymbol{Y}_{t-1})\right)e^{i\mu^\top\boldsymbol{Y}_{t-1}}\left(e^{i\nu^\top\boldsymbol{X}_{t-1}}-\widehat{\phi}(\nu| \boldsymbol{Y}_{t-1})\right),
\end{equation}
where $\left\{\xi_t\right\}_{t=1}^n$ is a sequence of i.i.d. random variables with zero mean, unit variance, and bounded support, independent of $\left\{\left(X_t, Y_t\right)^{\top}\right\}_{t=1}^n$. Note that if the residual $\left\{Y_t - \mathbb{E}(Y_t | \boldsymbol{W}_{t-1})\right\}$ is a martingale difference sequence, the choice of $\xi_t$ can fully characterize the covariance kernel of the process.

The asymptotic validity of the bootstrap procedure is formally established in the following theorem. Specifically, we show that the bootstrapped process $\widehat{S}_n^*(\mu, \nu)$ in \eqref{Snstar} converges weakly to the Gaussian process $S_{\infty}(\mu, \nu)$ in Theorem \ref{theoremnull}. 

\begin{theorem}
\label{theoremboo}
Suppose Assumptions \ref{betamixing}, \ref{convergencerate}, \ref{distribution} and \ref{local_alternative} hold. Then, under the null hypothesis in \eqref{drh}, the alternative hypothesis, or the sequence of local alternatives in \eqref{h1n},
$$
\widehat{S}_n^*(\cdot, \cdot) \underset{*}{\stackrel{P}{\rightarrow}} S_{\infty}(\cdot, \cdot),
$$
where $S_{\infty}(\cdot, \cdot)$ is the Gaussian process defined in Theorem \ref{theoremnull}. Here, $\underset{*}{\stackrel{P}{\rightarrow}}$ denotes the weak convergence in probability under the bootstrap law, as defined in \cite{gine1990bootstrapping}. Moreover, $\mathrm{KS}_n^* {\rightsquigarrow} \mathrm{KS}_{\infty}$, where $\mathrm{KS}_{\infty}$ is defined in Corollary \ref{conull}.
\end{theorem}

Theorem \ref{theoremboo} implies that the limiting behavior of $\widehat{S}_n(\mu, \nu)$ can be approximated by $\widehat{S}^*_n(\mu, \nu)$. Consequently, the bootstrap-assisted test statistic $\mathrm{KS}_n^*$ has the same convergence of $\mathrm{KS}_n$, constructed by substituting $\widehat{S}_n^*(\mu, \nu)$ for $\widehat{S}_n(\mu, \nu)$ in \eqref{KSn}. Therefore, given the significance level $\alpha$, the $p$-value of the test can be obtained by calculating the test statistic $\mathrm{KS}_n^*$ according to \textbf{Step 4} in Algorithm~\ref{algo}. In addition, $\mathrm{KS}_n^*$ is consistent against the alternative, and is able to detect local alternatives in \eqref{h1n}. See the detailed testing procedure in Algorithm~\ref{algo}.

\section{Simulation}
\label{simulations}

In this section, we conduct a set of Monte Carlo simulations to evaluate the finite-sample performance of the proposed test. 
Specifically, in Table \ref{DGPs}, the data generating process (DGP) S1 and S2 are considered to examine the size performance and P1-P4 are set for examining the power performance. Since our primary interest is testing the existence of Granger causality from $X_t$ to $Y_t$, we fix the dynamic of $X_t$ to AR($p$), while the connection between $\{X_{t-1}, \cdots, X_{t-p}\}$ and $Y_t$ covering different nonlinear patterns.
The innovations $\varepsilon_{1,t}$ and $\varepsilon_{2,t}$ are assumed to be i.i.d. $\mathcal{N}(0, 0.5)$ and mutually independent.

\begin{table*}[htb] \footnotesize
  \centering  
  \begin{threeparttable}  
  \caption{Data generating processes}
  \label{DGPs}   
  \setlength{\tabcolsep}{6.45mm}
    \begin{tabular}{cccccc}  
    \toprule[1.5pt]
    \multirow{2.5}{*}{DGPs}&  
    \multicolumn{2}{c}{Variables of interest}\cr  
    \cmidrule(lr){2-3}
    & $X_t$ & $Y_t$ \cr
    \midrule[1pt]
    DGP S1 &   $X_t =0.5 \mathop{\sum}\limits_{k=1}^{p} b_k{X}_{t-k} + \varepsilon_{1,t}$    &    $Y_t = 0.5 \mathop{\sum}\limits_{j=1}^{q} a_j{Y}_{t-j} + \varepsilon_{2,t}$   \cr
    DGP S2 &   $X_t =0.5 \mathop{\sum}\limits_{k=1}^{p} b_k{X}_{t-k} + \varepsilon_{1,t}$    &    $Y_t = a \mathop{\sum}\limits_{j=1}^{q} \mathrm{exp}\left(-0.5Y^2_{t-j}\right) + \varepsilon_{2,t}$   \cr
    DGP P1 &   $X_t =0.5 \mathop{\sum}\limits_{k=1}^{p} b_k{X}_{t-k} + \varepsilon_{1,t}$    &    $Y_t = 0.5 \mathop{\sum}\limits_{j=1}^{q} a_j{Y}_{t-j} +  \mathrm{sin} \left(\mathop{\sum}\limits_{k=1}^{p} c_k{X}_{t-k}\right) + \varepsilon_{2,t}$   \cr
    DGP P2 &   $X_t =0.5 \mathop{\sum}\limits_{k=1}^{p} b_k{X}_{t-k} + \varepsilon_{1,t}$    &   $Y_t = 0.5 \mathop{\sum}\limits_{j=1}^{q} a_j{Y}_{t-j} + 0.5c\mathop{\sum}\limits_{k=1}^{p} X_{t-k}^2  + \varepsilon_{2,t}$    \cr
    DGP P3 &   $X_t =0.5 \mathop{\sum}\limits_{k=1}^{p} b_k{X}_{t-k} + \varepsilon_{1,t}$    &   $Y_t = a \mathop{\sum}\limits_{j=1}^{q} \mathrm{exp}\left(-0.5Y^2_{t-j}\right) + c\mathop{\sum}\limits_{k=1}^{p} \mathrm{cos}\left(X_{t-k}\right) + \varepsilon_{2,t}$    \cr
    DGP P4 &   $X_t =0.5 \mathop{\sum}\limits_{k=1}^{p} b_k{X}_{t-k} + \varepsilon_{1,t}$    &   $Y_t = a_0 \mathop{\sum}\limits_{j=1}^{q} \left(X_{t-j}Y_{t-j}\right) + \varepsilon_{2,t}$    \cr
    \bottomrule[1.5pt]
    \end{tabular} 
    \end{threeparttable}  
\end{table*}


Henceforth, the terms lag order and lag length are used interchangeably and referred to simply as $\mathrm{lag}$. $\mathrm{Lag}$ specifies the number of lags of $Y_t$ or $X_t$, and thus defines the dimensionality of the input to the deep learning model, and in our setting it satisfies that $\mathrm{lag} = p = q$. For each DGP, we examine five scenarios corresponding to $\mathrm{lag} = 1,2,3,4$ and $5$. The following is the specification of the parameters, and we summarize the settings in Table \ref{DGPpara}.
For $\mathrm{lag}=1$, we set $a_1=1$, $b_1=-1$, $c_1=-1$, $a=0.5$, $c=1$, and $ a_0=0.5$. For $\mathrm{lag}=2$, we set $(a_1,a_2)^\top=(0.5,-0.5)^\top$, $(b_1,b_2)^\top=(-0.5,0.5)^\top$, $(c_1,c_2)^\top=(-0.5,0.5)^\top$, $a=0.25$, $c=0.6$, and $ a_0=0.4$. For $\mathrm{lag}=3$, we set $(a_1,a_2,a_3)^\top=(0.5,-0.5,0.5)^\top$, $(b_1,b_2,b_3)^\top=(-0.5,0.5,0.5)^\top$, $(c_1,c_2,c_3)^\top=(-0.5,0.5,-0.5)^\top$, $a=0.25$, $c=0.5$, and $ a_0=1/3$.
For $\mathrm{lag}=4$, we set $(a_1,a_2,a_3,a_4)^\top=(0.25,-0.25,0.25,0.25)^\top$, $(b_1,b_2,b_3,b_4)^\top=(-0.25,0.25,0.25,-0.25)^\top$, $(c_1,c_2,c_3,c_4)^\top=(-0.25,0.25,-0.25,0.25)^\top$, $a=0.125$, $c=0.5$, and $ a_0=1/3$. For $\mathrm{lag}=5$, we set $(a_1,a_2,a_3,a_4,a_5)^\top=(0.25,-0.25,0.25,0.25,-0.25)^\top$, $(b_1,b_2,b_3,b_4,b_5)^\top=(-0.25,0.25,0.25,-0.25,0.25)^\top$, $(c_1,c_2,c_3,c_4,c_5)^\top=(-0.25,0.25,-0.25,0.25,-0.25)^\top$, $a=0.125$, $c=0.5$, and $ a_0=1/3$.
These parameters are designed to constrain the summation in $Y_t$ across multiple lags to prevent it from diverging during data generation, although they may not produce the best performance for our test.

\begin{table*}[htb] \footnotesize
  \centering  
  \begin{threeparttable}  
  \caption{Parameter settings} 
  \label{DGPpara}   
  \setlength{\tabcolsep}{0.61mm}
    \begin{tabular}{ccccccc}  
    \toprule[1.5pt]
    \multirow{2.5}{*}{Lag} & \multicolumn{6}{c}{Parameters in DGPs} \cr
     \cmidrule{2-7}
    & S1 & S2 & P1 & P2 & P3 & P4 \cr
    \cmidrule[1pt](lr){1-7}
    \multirow{3}{*}{$\mathrm{lag}=1$} & \multirow{3}{*}{\begin{tabular}[c]{@{}c@{}} $a_1=1$ \\ $b_1=-1$ \end{tabular}} & \multirow{3}{*}{\begin{tabular}[c]{@{}c@{}} $a=0.5$ \\ $b_1=-1$ \end{tabular}} & $a_1=1$ & \multirow{3}{*}{\begin{tabular}[c]{@{}c@{}} $a_1=1$ \\ $b_1=-1$ \\ $c=1$ \end{tabular}} & \multirow{3}{*}{\begin{tabular}[c]{@{}c@{}} $a=0.5$ \\ $b_1=-1$ \\ $c=1$ \end{tabular}} & \multirow{3}{*}{\begin{tabular}[c]{@{}c@{}} $a_0=0.5$ \\ $b_1=-1$ \end{tabular}} \cr 
    &  &  & $b_1=-1$ &  &  & \cr 
    &  &  & $c_1=-1$ &  &  & \cr 
    \cmidrule(lr){1-7}
    \multirow{6}{*}{$\mathrm{lag}=2$} & \multirow{6}{*}{\begin{tabular}[c]{@{}c@{}} $a_1=0.5$ \\ $a_2=-0.5$ \\ $b_1=-0.5$ \\ $b_2=0.5$ \end{tabular}} & \multirow{6}{*}{\begin{tabular}[c]{@{}c@{}} $a=0.25$ \\ $b_1=-0.5$ \\ $b_2=0.5$ \end{tabular}} & $a_1=0.5$ & \multirow{6}{*}{\begin{tabular}[c]{@{}c@{}} $a_1=0.5$ \\ $a_2=-0.5$ \\ $b_1=-0.5$ \\ $b_2=0.5$ \\ $c=0.6$ \end{tabular}} & \multirow{6}{*}{\begin{tabular}[c]{@{}c@{}} $a=0.25$ \\ $b_1=-0.5$ \\ $b_2=0.5$ \\ $c=0.6$ \end{tabular}} & \multirow{6}{*}{\begin{tabular}[c]{@{}c@{}} $a_0=0.4$ \\ $b_1=-0.5$ \\ $b_2=0.5$ \end{tabular}} \cr 
    &  &  & $a_2=-0.5$ &  &  & \cr 
    &  &  & $b_1=-0.5$ &  &  & \cr 
    &  &  & $b_2=0.5$ &  &  & \cr 
    &  &  & $c_1=-0.5$ &  &  & \cr 
    &  &  & $c_2=0.5$ &  &  & \cr  
    \cmidrule(lr){1-7}
    \multirow{6}{*}{$\mathrm{lag}=3$} & \multirow{6}{*}{\begin{tabular}[c]{@{}c@{}} $a_1=a_3=0.5$ \\ $a_2=-0.5$ \\ $b_1=-0.5$ \\ $b_2=b_3=0.5$ \end{tabular}} & \multirow{6}{*}{\begin{tabular}[c]{@{}c@{}} $a=0.25$ \\ $b_1=-0.5$ \\ $b_2=b_3=0.5$ \end{tabular}} & $a_1=a_3=0.5$ & \multirow{6}{*}{\begin{tabular}[c]{@{}c@{}} $a_1=a_3=0.5$ \\ $a_2=-0.5$ \\ $b_1=-0.5$ \\ $b_2=b_3=0.5$ \\ $c=0.5$ \end{tabular}} & \multirow{6}{*}{\begin{tabular}[c]{@{}c@{}} $a=0.25$ \\ $b_1=-0.5$ \\ $b_2=b_3=0.5$ \\ $c=0.5$ \end{tabular}} & \multirow{6}{*}{\begin{tabular}[c]{@{}c@{}} $a_0=1/3$ \\ $b_1=-0.5$ \\ $b_2=b_3=0.5$ \end{tabular}} \cr 
    &  &  & $a_2=-0.5$ &  &  & \cr 
    &  &  & $b_1=-0.5$ &  &  & \cr 
    &  &  & $b_2=b_3=0.5$ &  &  & \cr 
    &  &  & $c_1=c_3=-0.5$ &  &  & \cr 
    &  &  & $c_2=0.5$ &  &  & \cr 
    \cmidrule(lr){1-7}
    \multirow{6}{*}{$\mathrm{lag}=4$} & \multirow{6}{*}{\begin{tabular}[c]{@{}c@{}} $a_1=a_3=a_4=0.25$ \\ $a_2=-0.25$ \\ $b_1=b_4=-0.25$ \\ $b_2=b_3=0.25$ \end{tabular}} & \multirow{6}{*}{\begin{tabular}[c]{@{}c@{}} $a=0.125$ \\ $b_1=b_4=-0.25$ \\ $b_2=b_3=0.25$ \end{tabular}} & $a_1=a_3=a_4=0.25$ & \multirow{6}{*}{\begin{tabular}[c]{@{}c@{}} $a_1=a_3=a_4=0.25$ \\ $a_2=-0.25$ \\ $b_1=b_4=-0.25$ \\ $b_2=b_3=0.25$ \\ $c=0.5$ \end{tabular}} & \multirow{6}{*}{\begin{tabular}[c]{@{}c@{}} $a=0.125$ \\ $b_1=b_4=-0.25$ \\ $b_2=b_3=0.25$ \\ $c=0.5$ \end{tabular}} & \multirow{6}{*}{\begin{tabular}[c]{@{}c@{}} $a_0=1/3$ \\ $b_1=b_4=-0.25$ \\ $b_2=b_3=0.25$ \end{tabular}} \cr 
    &  &  & $a_2=-0.25$ &  &  & \cr 
    &  &  & $b_1=b_4=-0.25$ &  &  & \cr 
    &  &  & $b_2=b_3=0.25$ &  &  & \cr 
    &  &  & $c_1=c_3=-0.25$ &  &  & \cr 
    &  &  & $c_2=c_4=0.25$ &  &  & \cr 
    \cmidrule(lr){1-7}
    \multirow{6}{*}{$\mathrm{lag}=5$} & \multirow{6}{*}{\begin{tabular}[c]{@{}c@{}} $a_1=a_3=a_4=0.25$ \\ $a_2=a_5=-0.25$ \\ $b_1=b_4=-0.25$ \\ $b_2=b_3=b_5=0.25$ \end{tabular}} & \multirow{6}{*}{\begin{tabular}[c]{@{}c@{}} $a=0.125$ \\ $b_1=b_4=-0.25$ \\ $b_2=b_3=b_5=0.25$ \end{tabular}} & $a_1=a_3=a_4=0.25$ & \multirow{6}{*}{\begin{tabular}[c]{@{}c@{}} $a_1=a_3=a_4=0.25$ \\ $a_2=a_5=-0.25$ \\ $b_1=b_4=-0.25$ \\ $b_2=b_3=b_5=0.25$ \\ $c=0.5$ \end{tabular}} & \multirow{6}{*}{\begin{tabular}[c]{@{}c@{}} $a=0.125$ \\ $b_1=b_4=-0.25$ \\ $b_2=b_3=b_5=0.25$ \\ $c=0.5$ \end{tabular}} & \multirow{6}{*}{\begin{tabular}[c]{@{}c@{}} $a_0=1/3$ \\ $b_1=b_4=-0.25$ \\ $b_2=b_3=b_5=0.25$ \end{tabular}} \cr 
    &  &  & $a_2=a_5=-0.25$ &  &  & \cr 
    &  &  & $b_1=b_4=-0.25$ &  &  & \cr 
    &  &  & $b_2=b_3=b_5=0.25$ &  &  & \cr 
    &  &  & $c_1=c_3=c_5=-0.25$ &  &  & \cr 
    &  &  & $c_2=c_4=0.25$ &  &  & \cr 
    \bottomrule[1.5pt]
    \end{tabular} 
    \end{threeparttable}  
\end{table*}

For hyperparameters, we select the number of mixture components $G$ via cross-validation, as it affects empirical performance to a certain extent. A small value of $G$ can introduce a large bias in the fitted MDN model, leading to inflated type I error rates. In contrast, a large $G$ can cause overfitting, resulting in increased variability of the test statistic. In general, the setting $G=10$ is suitable for most scenarios. Regarding $L$, that is the number of $(\mu,\nu)$ pairs, a higher value of $L$ generally increases the power of the test, but at the cost of higher computational complexity. To balance power performance and efficiency, we fix $L = 20$. For the number of pseudo-samples $M$, we find that the test is relatively insensitive to this parameter, provided $M$ within a reasonable range. Hence, we fix $M = 20$ for all numerical studies.

For both MLP and MDN models, we fix the depth of neural networks $L_n$ to be $L_n = 1$. The width, that is the number of nodes $H_n$ per hidden layer, is varied with the lag as $H_n=5\cdot\mathrm{lag}$. This choice provides a balance between network simplicity and simulation performance. For the MLP, the loss function can be chosen as $L_2$-loss or smooth $L_1$-loss, while the parameters are estimated using maximum likelihood for the MDN. 

Let $n_0=500$. For each DGP, we generate $n_0+n$ observations and discard the initial $n_0$ observations to mitigate the effects of the initial values. The simulations are based on $1000$ Monte Carlo replications, and the bootstrap $p$-values are computed using $B = 1000$ bootstrap replications. Three sample sizes are considered, $n = 500$, 1000, and 2000. We report results for a nominal significance level of $5\%$, with results for other levels available upon request.

Since there is no causal relationship between $X_t$ and $Y_t$ in DGP S1 and S2, the null hypothesis of non-Granger causality should be maintained. This experiment visually demonstrates the decent size properties of our test as the lags of $\boldsymbol{X}_{t-1}$ and $\boldsymbol{Y}_{t-1}$ vary. Each entry shows the empirical rejection rate of the null hypothesis of non-Granger causality. In the table, we denote the doubly robust Granger causality test proposed in this paper as DRGC. For comparison, the nonparametric causality test proposed by \cite{nishiyama2011consistent} is denoted as NHKJ. The results are summarized in Table \ref{DGPsize}.

\begin{table*}[htb] \footnotesize
  \centering  
  \begin{threeparttable}  
  \caption{Empirical sizes under varying lags} 
  \label{DGPsize}   
  \setlength{\tabcolsep}{5.67mm}
    \begin{tabular}{ccccccc}  
    \toprule[1.5pt]
    \multirow{2.5}{*}{Lag}&\multicolumn{3}{c}{DGP S1} & \multicolumn{3}{c}{DGP S2} \cr 
    \cmidrule(lr){2-4} \cmidrule(lr){5-7}
                      & Sample Size & DRGC & NHKJ & Sample Size & DRGC & NHKJ \cr
    \cmidrule[1pt](l){1-1}\cmidrule[1pt]{2-4} \cmidrule[1pt](r){5-7}
    \multirow{3}{*}{$\mathrm{lag}=1$}&$n=500$ & 0.051 & 0.039 & $n=500$ & 0.046 & 0.036 \cr 
                      &$n=1000$& 0.051 & 0.051 & $n=1000$& 0.044 & 0.068 \cr 
                      &$n=2000$& 0.057 & 0.046 & $n=2000$& 0.056 & 0.178 \cr 
                      \cmidrule(l){1-1}\cmidrule{2-4} \cmidrule(r){5-7}
    \multirow{3}{*}{$\mathrm{lag}=2$}&$n=500$ & 0.056 & 0.016 & $n=500$ & 0.049 & 0.006 \cr 
                      &$n=1000$& 0.049 & 0.019 & $n=1000$& 0.038 & 0.011 \cr 
                      &$n=2000$& 0.055 & 0.026 & $n=2000$& 0.051 & 0.023 \cr 
                      \cmidrule(l){1-1}\cmidrule{2-4} \cmidrule(r){5-7}
    \multirow{3}{*}{$\mathrm{lag}=3$}&$n=500$ & 0.046 & 0.008 & $n=500$ & 0.045 & 0.003 \cr 
                      &$n=1000$& 0.059 & 0.025 & $n=1000$& 0.051 & 0.007 \cr 
                      &$n=2000$& 0.052 & 0.043 & $n=2000$& 0.045 & 0.010 \cr 
                      \cmidrule(l){1-1}\cmidrule{2-4} \cmidrule(r){5-7}
    \multirow{3}{*}{$\mathrm{lag}=4$}&$n=500$ & 0.057 & 0.006 & $n=500$ & 0.039 & 0.005 \cr 
                      &$n=1000$& 0.042 & 0.008 & $n=1000$& 0.055 & 0.006 \cr 
                      &$n=2000$& 0.047 & 0.012 & $n=2000$& 0.055 & 0.011 \cr 
                      \cmidrule(l){1-1}\cmidrule{2-4} \cmidrule(r){5-7}
    \multirow{3}{*}{$\mathrm{lag}=5$}&$n=500$ & 0.050 & 0.003 & $n=500$ & 0.051 & 0.005 \cr 
                      &$n=1000$& 0.049 & 0.009 & $n=1000$& 0.049 & 0.004 \cr 
                      &$n=2000$& 0.056 & 0.007 & $n=2000$& 0.050 & 0.014 \cr 
    \bottomrule[1.5pt]
    \end{tabular}  
    \end{threeparttable}  
\end{table*}

It is worth noting that the results of size are reasonable when the sample sizes are increasing, such as $n=1000$ and $n=2000$. However, if hastily integrate deep neural network estimators \textemdash specifically, by using the hypothesis in \eqref{uncondi} to construct a Granger causality test \textemdash this approach fails to control the type I error, especially when the sample size increases. For example, in DGP S1, the size is 0.151 for $n=1000$ and 0.321 for $n=2000$ under $\mathrm{lag}=5$. This phenomenon highlights the necessity for the doubly robust Granger causality test, which has facilitated the practical integration of deep learning into statistical testing methods. Furthermore, as observed, dimension does not significantly affect the size property under the proposed test. 
We also employ the test proposed by \cite{nishiyama2011consistent}. Since the dimensionality is up to 5, we adopt a fourth-order Gaussian kernel function for NHKJ to align with the theoretical analysis. To ensure fairness, we set different bandwidths in S1 and S2 across varying lags for NHKJ. For DGP S1, set bandwidth $h=2.5\times n^{-0.15}$ when $\mathrm{lag}=1$, and $h=3\times n^{-0.15}$ when $\mathrm{lag}=2,3$, while $h=3.5\times n^{-0.15}$ when $\mathrm{lag}=4,5$. Similarly, for DGP S2, set bandwidth $h=3\times T^{-0.15}$ when $\mathrm{lag}=1$, and $h=3.5\times n^{-0.15}$ when $\mathrm{lag}=2,3$, while $h=4\times n^{-0.15}$ when $\mathrm{lag}=4,5$. According to the results in Table \ref{DGPsize}, NHKJ struggles to control type I error at the nominal level and exhibits a trend of increasing type I error as the sample size grows. This may be attributed to NHKJ’s sensitivity to bandwidth selection. We do not choose a sufficiently large bandwidth to control the type I error at the nominal significance level because doing so would require an excessively large value. For example, in the case of S1 with $\mathrm{lag}=5$, the bandwidth should be set to $4\times n^{-0.15}$, whereas in the case of S2 with $\mathrm{lag}=5$, the bandwidth is required to exceed $8\times n^{-0.15}$.

Then, we investigate the power performance of the proposed test. Specifically, DGP P1, P2, P3 and P4 are designed to evaluate the power while the lag parameter $\mathrm{lag}$ varies from 1 to 5, examining the impact of increasing lag orders on power performance.
DGP P1, P2, and P3 consider an additive structure for $Y_t$, where $Y_t$ is modeled as the sum of a pure $\boldsymbol{Y}_{t-1}$ component and a pure $\boldsymbol{X}_{t-1}$ component. Particularly, P1 and P2 are designed with a linear combination of $\boldsymbol{Y}_{t-1}$, while P3 introduces a more complex nonlinear structure for $\boldsymbol{Y}_{t-1}$. Regarding the $\boldsymbol{X}_{t-1}$ component in $Y_t$, P1 incorporates a linear summation of $\boldsymbol{X}_{t-1}$, which is then transformed by a nonlinear function, that is $\sin$. In contrast, P2 and P3 apply a nonlinear transformation to $\{X_{t-1},\cdots,X_{t-p}\}$ individually before summation. 
DGP P4, unlike the other three settings, adopts a more hybrid structure. It includes the product of $\{X_{t-i},Y_{t-i}\}_{i=1}^q$, rather than treating $\boldsymbol{X}_{t-1}$ and $\boldsymbol{Y}_{t-1}$ as separate components.

Since there is an actual causality between $X_t$ and $Y_t$, the alternative hypothesis holds. This experiment visually shows the changes of power properties of the proposed test as the lag orders of $\boldsymbol{X}_{t-1}$ and $\boldsymbol{Y}_{t-1}$ vary. Each entry in the tables presents the empirical rejection rate of the null hypothesis of non-causality. The results are summarized in Table \ref{DGPpower}.

As observed, our method performs strongly when the dimensionality ranges from one to three, with power approaching one. While performance declines with increasing dimensionality, the test remains effective in cases with lag orders of four and five, which are generally considered difficult for nonparametric methods. This indicates a potential improvement in large-lag Granger causality testing. For NHKJ, note that DGP P1 and P2 correspond to DGP S1, while DGP P3 corresponds to DGP S2. Therefore, the bandwidth settings are aligned with those used in S1 and S2, respectively. As shown, the power of NHKJ is significantly affected as the lag order increases—even at $\mathrm{lag} = 2$, it exhibits noticeably lower power compared to the doubly robust test. This underscores the advantage of our proposed method in testing Granger causality across multiple lags.

\begin{table*}[htb] \footnotesize
  \centering  
  \begin{threeparttable}  
  \caption{Empirical powers under varying lags} 
  \label{DGPpower}   
  \setlength{\tabcolsep}{5.67mm}
    \begin{tabular}{ccccccc}  
    \toprule[1.5pt]
    \multirow{2.5}{*}{Lag}&\multicolumn{3}{c}{DGP P1} & \multicolumn{3}{c}{DGP P2} \cr 
    \cmidrule(lr){2-4} \cmidrule(lr){5-7}
                      & Sample Size & DRGC & NHKJ & Sample Size & DRGC & NHKJ \cr
    \cmidrule[1pt](l){1-1}\cmidrule[1pt]{2-4} \cmidrule[1pt](r){5-7}
    \multirow{3}{*}{$\mathrm{lag}=1$}&$n=500$ & 1.000 & 1.000 & $n=500$ & 0.996 & 1.000 \cr 
                      &$n=1000$& 1.000 & 1.000 & $n=1000$& 1.000 & 1.000 \cr 
                      &$n=2000$& 1.000 & 1.000 & $n=2000$& 1.000 & 1.000 \cr 
                      \cmidrule(l){1-1}\cmidrule{2-4} \cmidrule(r){5-7}
    \multirow{3}{*}{$\mathrm{lag}=2$}&$n=500$ & 1.000 & 0.994 & $n=500$ & 0.973 & 0.459 \cr 
                      &$n=1000$& 1.000 & 1.000 & $n=1000$& 1.000 & 0.991 \cr 
                      &$n=2000$& 1.000 & 1.000 & $n=2000$& 1.000 & 1.000 \cr 
                      \cmidrule(l){1-1}\cmidrule{2-4} \cmidrule(r){5-7}
    \multirow{3}{*}{$\mathrm{lag}=3$}&$n=500$ & 1.000 & 0.991 & $n=500$ & 0.898 & 0.369 \cr 
                      &$n=1000$& 1.000 & 1.000 & $n=1000$& 1.000 & 0.929 \cr 
                      &$n=2000$& 1.000 & 1.000 & $n=2000$& 1.000 & 1.000 \cr 
                      \cmidrule(l){1-1}\cmidrule{2-4} \cmidrule(r){5-7}
    \multirow{3}{*}{$\mathrm{lag}=4$}&$n=500$ & 0.912 & 0.546 & $n=500$ & 0.628 & 0.188 \cr 
                      &$n=1000$& 1.000 & 0.974 & $n=1000$& 0.956 & 0.739 \cr 
                      &$n=2000$& 1.000 & 1.000 & $n=2000$& 1.000 & 0.994 \cr 
                      \cmidrule(l){1-1}\cmidrule{2-4} \cmidrule(r){5-7}
    \multirow{3}{*}{$\mathrm{lag}=5$}&$n=500$ & 0.870 & 0.233 & $n=500$ & 0.546 & 0.052 \cr 
                      &$n=1000$& 0.991 & 0.694 & $n=1000$& 0.905 & 0.454 \cr 
                      &$n=2000$& 1.000 & 0.979 & $n=2000$& 0.998 & 0.957 \cr 
    \midrule[1pt]
    \multirow{2.5}{*}{Lag}&\multicolumn{3}{c}{DGP P3} & \multicolumn{3}{c}{DGP P4} \cr 
    \cmidrule(lr){2-4} \cmidrule(lr){5-7}
                      & Sample Size & DRGC & NHKJ & Sample Size & DRGC & NHKJ \cr
    \cmidrule[1pt](l){1-1}\cmidrule[1pt]{2-4} \cmidrule[1pt](r){5-7}
    \multirow{3}{*}{$\mathrm{lag}=1$}&$n=500$ & 1.000 & 1.000 & $n=500$ & 0.990 & / \cr 
                      &$n=1000$& 1.000 & 1.000 & $n=1000$& 1.000 & / \cr 
                      &$n=2000$& 1.000 & 1.000 & $n=2000$& 1.000 & / \cr 
                      \cmidrule(l){1-1}\cmidrule{2-4} \cmidrule(r){5-7}
    \multirow{3}{*}{$\mathrm{lag}=2$}&$n=500$ & 0.977 & 0.315 & $n=500$ & 0.959 & / \cr 
                      &$n=1000$& 1.000 & 0.930 & $n=1000$& 1.000 & / \cr 
                      &$n=2000$& 1.000 & 1.000 & $n=2000$& 1.000 & / \cr 
                      \cmidrule(l){1-1}\cmidrule{2-4} \cmidrule(r){5-7}
    \multirow{3}{*}{$\mathrm{lag}=3$}&$n=500$ & 0.954 & 0.162 & $n=500$ & 0.886 & / \cr 
                      &$n=1000$& 1.000 & 0.779 & $n=1000$& 1.000 & / \cr 
                      &$n=2000$& 1.000 & 1.000 & $n=2000$& 1.000 & / \cr 
                      \cmidrule(l){1-1}\cmidrule{2-4} \cmidrule(r){5-7}
    \multirow{3}{*}{$\mathrm{lag}=4$}&$n=500$ & 0.429 & 0.060 & $n=500$ & 0.663 & / \cr 
                      &$n=1000$& 0.776 & 0.461 & $n=1000$& 0.965 & / \cr 
                      &$n=2000$& 0.981 & 0.969 & $n=2000$& 1.000 & / \cr 
                      \cmidrule(l){1-1}\cmidrule{2-4} \cmidrule(r){5-7}
    \multirow{3}{*}{$\mathrm{lag}=5$}&$n=500$ & 0.407 & 0.087 & $n=500$ & 0.600 & / \cr 
                      &$n=1000$& 0.747 & 0.437 & $n=1000$& 0.917 & / \cr 
                      &$n=2000$& 0.974 & 0.948 & $n=2000$& 1.000 & / \cr 
    \bottomrule[1.5pt]
    \end{tabular}  
    \end{threeparttable}  
\end{table*}

\section{Real data application}
\label{real data}
In this section, we examine the presence of Granger causality between stock prices and trading volumes using empirical data. It is well established in the finance literature that stock price fluctuations and trading volumes are closely related across adjacent periods \textemdash a  phenomenon extensively documented in both economics and financial research (see \citeauthor{karpoff1987relation}, \citeyear{karpoff1987relation}; \citeauthor{gallant1992stock}, \citeyear{gallant1992stock}; \citeauthor{chen2001forecasting}, \citeyear{chen2001forecasting}). This price–volume relationship is widely regarded as an important indicator for anticipating future market trends. In general, trading volumes tend to rise during periods of substantial price movements and decline when price changes are less pronounced. Conversely, periods of high trading volume are often followed by significant price fluctuations.

We use daily prices and volumes data from three renowned stock market indices representing different economies: the S\&P 500 Index (SPX 500), the CSI 300 Index (CSI 300), and the Nikkei 225 Index (NI 225). The sample period spans five years, from September 27, 2019 to September 26, 2024, yielding a total sample size of $T = 1257$. To ensure stationarity, both prices and volumes are transformed into percentage changes. For volumes, the percentage change is divided by 10 to conform model training and achieve a scale comparable to that of stock prices. We denote these percentage-change variables as $P_t$ for prices and $V_t$ for volumes. 

To investigate the stability and consistency of Granger causality patterns, we conduct analyses using overlapping three-year sub-samples for each index. Specifically, we examine Granger causality using data from September 27, 2019 to September 26, 2022, from September 27, 2020 to September 26, 2023, and from September 27, 2021 to September 26, 2024, hereafter referred to as 2019-2022, 2020-2023, and 2021-2024, respectively. The sample size for each three-year period is approximately $n = 750$.

The primary objective is to investigate whether there exists causality in the mean between changes in trading volumes and stock prices. Specifically, we aim to detect bidirectional Granger causality at lag orders ranging from 1 to 10 using the proposed test. To test whether price changes Granger-cause volume changes, we treat $P_t$ as $X_t$ and $V_t$ as $Y_t$. Conversely, testing whether volume changes Granger-cause price changes reverses these roles. The parameter settings are mostly consistent with those in the simulation studies, except for setting the lag length from 1 to 10, as the computational cost in this application is significantly lower than in the simulations. The upper 5\% critical value is employed as the threshold for hypothesis testing.

Table \ref{pricevolume} presents a short summary of the numerical results. As shown in the table, when examining Granger causality from $P_t$ to $V_t$, the null hypothesis of non-causality is rejected for the SPX 500 and CSI 300 indices in most periods. However, no evidence of causality is found for the NI 225 index across any of the three periods. Conversely, when considering Granger causality from $V_t$ to $P_t$, all three indices seldom exhibit causality over time.

In addition, we provide the test results specific to different lag orders in Table \ref{pricevolumedetail}.
For the SPX 500 index, significant Granger causality from $P_t$ to $V_t$ is observed at all lag orders across the 2019-2022 and 2021-2024 periods. In contrast, during 2020–2023, causality emerges only from lag order 4 onward. In the reverse direction, no evidence of causality from $V_t$ to $P_t$ is found at any lag order. For the CSI 300 index, $P_t$ Granger-causes $V_t$ in both the 2020–2023 and 2021–2024 periods, though the relevant lag orders differ. Specifically, in 2020–2023, causality first appears at lag order 7, while in 2021–2024 it spans lag orders 2 through 10. Conversely, causality from $V_t$ to $P_t$ is detected only in 2021–2024, also across lag orders 2 through 10. For the NI 225 index, no evidence of Granger causality is found in either direction between $P_t$ and $V_t$ during any period.

The numerical results yield several noteworthy conclusions. For each index, we observe either a consistent or a shifting pattern of Granger causality in mean from stock prices to trading volumes. The characteristics differ across indices: for the SPX 500, there appears to be persistent causality from prices to volumes; for the CSI 300, the Granger causality appears to evolve over time, becoming more pronounced in later periods. and for the Nikkei 225, no evidence of causality is found across all periods. Notably, such causality does not necessarily emerge immediately but often manifests over multiple lag orders. 
Our proposed method is particularly effective in capturing these patterns, especially when causality extends across several lags. An additional advantage of DRGCT is its robustness in practical applications: since the true lag order is typically unknown in real-world data, selecting the appropriate lag length can be challenging. By applying our test, researchers can set the lag order to a sufficiently large value, thereby encompassing all potential causal lags.

These findings may also reflect general market behavior. A temporary price change does not necessarily exert an immediate impact on trading activity. By contrast, sustained price trends are more likely to affect trading, as investors tend to base their decisions on cumulative patterns rather than isolated daily fluctuations. This interpretation is consistent with \cite{copeland1976model}, whose sequential information arrival model shows that trading volume responds more strongly to the cumulative impact of information arrivals rather than to isolated shocks, and also relates to \cite{de1990noise}, who demonstrate how noise traders can cause persistent price deviations and market anomalies.

In the reverse direction, trading volumes occasionally serve as a causal factor for stock prices, though such relationships appear far less consistent. It is worth noting that our empirical analysis is restricted to daily data with lag orders up to 10. Thus, causal relationships at longer horizons cannot be ruled out. Moreover, our framework considers only cases where both series share the same lag order, suggesting that more nuanced causal structures may exist and warrant further exploration.

\begin{table*}[htb] \footnotesize
  \centering  
  \begin{threeparttable}  
  \caption{Price-volume Granger causality detection} 
  \label{pricevolume}   
  \setlength{\tabcolsep}{6.05mm}
    \begin{tabular}{ccccccc}  
    \toprule[1.5pt]
    \multirow{2.5}{*}{Causality Direction} & \multicolumn{6}{c}{Index Category} \cr
    \cmidrule(lr){2-7}
                      &\multicolumn{2}{c}{SPX 500} & \multicolumn{2}{c}{CSI 300} & \multicolumn{2}{c}{NI 225} \cr 
    \cmidrule[1pt](l){1-1}\cmidrule[1pt]{2-3} \cmidrule[1pt]{4-5} \cmidrule[1pt](r){6-7}
    \multirow{3}{*}{$P_t\rightarrow V_t$} &2019-2022& \ding{51} &2019-2022& \ding{55} &2019-2022& \ding{55} \cr 
                                          &2020-2023& \ding{51} &2020-2023& \ding{51} &2020-2023& \ding{55} \cr 
                                          &2021-2024& \ding{51} &2021-2024& \ding{51} &2021-2024& \ding{55} \cr 
                      \cmidrule(l){2-3} \cmidrule{4-5} \cmidrule(r){6-7}
    \multirow{3}{*}{$V_t\rightarrow P_t$} &2019-2022& \ding{55} &2019-2022& \ding{55} &2019-2022& \ding{55} \cr 
                                          &2020-2023& \ding{55} &2020-2023& \ding{55} &2020-2023& \ding{55} \cr 
                                          &2021-2024& \ding{55} &2021-2024& \ding{51} &2021-2024& \ding{55} \cr 
    \bottomrule[1.5pt]
    \end{tabular}  
    \end{threeparttable}  
\end{table*}

\newpage
\begin{table*}[htb] \footnotesize
  \centering  
  \begin{threeparttable}  
  \caption{Price-volume Granger causality under specific lag orders} 
  \label{pricevolumedetail}   
  \setlength{\tabcolsep}{3.17mm}
    \begin{tabular}{ccccccccccccc}  
    \toprule[1.5pt]
    \multirow{2.5}{*}{Causality Direction} & \multirow{2.5}{*}{Index} & \multirow{2.5}{*}{Period} & \multicolumn{10}{c}{Lag Order} \cr
    \cmidrule(lr){4-13}
                      &&&{1} & {2} & {3} &{4} & {5} & {6} &{7} & {8} & {9} & {10} \cr 
    \midrule[1pt]
    \multirow{9}{*}{$P_t\rightarrow V_t$}& \multirow{3}{*}{SPX 500}   & 2019-2022 & \ding{51} & \ding{51} & \ding{51} & \ding{51} & \ding{51} & \ding{51} & \ding{51} & \ding{51} & \ding{51} & \ding{51} \cr 
                                          &                            & 2020-2023 & \ding{55} & \ding{55} & \ding{55} & \ding{51} & \ding{51} & \ding{51} & \ding{51} & \ding{51} & \ding{51} & \ding{51} \cr 
                                          &                            & 2021-2024 & \ding{51} & \ding{51} & \ding{51} & \ding{51} & \ding{51} & \ding{51} & \ding{51} & \ding{51} & \ding{51} & \ding{51} \cr 
    \cmidrule(lr){2-13}
                                          & \multirow{3}{*}{CSI 300}   & 2019-2022 & \ding{55} & \ding{55} & \ding{55} & \ding{55} & \ding{55} & \ding{55} & \ding{55} & \ding{55} & \ding{55} & \ding{55} \cr 
                                          &                            & 2020-2023 & \ding{55} & \ding{55} & \ding{55} & \ding{55} & \ding{55} & \ding{55} & \ding{51} & \ding{51} & \ding{51} & \ding{51} \cr 
                                          &                            & 2021-2024 & \ding{55} & \ding{51} & \ding{51} & \ding{51} & \ding{51} & \ding{51} & \ding{51} & \ding{51} & \ding{51} & \ding{51} \cr 
    \cmidrule(lr){2-13}
                                          & \multirow{3}{*}{NI 225}    & 2019-2022 & \ding{55} & \ding{55} & \ding{55} & \ding{55} & \ding{55} & \ding{55} & \ding{55} & \ding{55} & \ding{55} & \ding{55} \cr 
                                          &                            & 2020-2023 & \ding{55} & \ding{55} & \ding{55} & \ding{55} & \ding{55} & \ding{55} & \ding{55} & \ding{55} & \ding{55} & \ding{55} \cr 
                                          &                            & 2021-2024 & \ding{55} & \ding{55} & \ding{55} & \ding{55} & \ding{55} & \ding{55} & \ding{55} & \ding{55} & \ding{55} & \ding{55} \cr 
    \midrule[1pt]
    \multirow{2.5}{*}{Causality Direction} & \multirow{2.5}{*}{Index} & \multirow{2.5}{*}{Period} & \multicolumn{10}{c}{Lag Order} \cr
    \cmidrule(lr){4-13}
                      &&&{1} & {2} & {3} &{4} & {5} & {6} &{7} & {8} & {9} & {10} \cr 
    \midrule[1pt]
    \multirow{9}{*}{$V_t\rightarrow P_t$}& \multirow{3}{*}{SPX 500}   & 2019-2022 & \ding{55} & \ding{55} & \ding{55} & \ding{55} & \ding{55} & \ding{55} & \ding{55} & \ding{55} & \ding{55} & \ding{55} \cr 
                                          &                            & 2020-2023 & \ding{55} & \ding{55} & \ding{55} & \ding{55} & \ding{55} & \ding{55} & \ding{55} & \ding{55} & \ding{55} & \ding{55} \cr 
                                          &                            & 2021-2024 & \ding{55} & \ding{55} & \ding{55} & \ding{55} & \ding{55} & \ding{55} & \ding{55} & \ding{55} & \ding{55} & \ding{55} \cr 
    \cmidrule(lr){2-13}
                                          & \multirow{3}{*}{CSI 300}   & 2019-2022 & \ding{55} & \ding{55} & \ding{55} & \ding{55} & \ding{55} & \ding{55} & \ding{55} & \ding{55} & \ding{55} & \ding{55} \cr 
                                          &                            & 2020-2023 & \ding{55} & \ding{55} & \ding{55} & \ding{55} & \ding{55} & \ding{55} & \ding{55} & \ding{55} & \ding{55} & \ding{55} \cr 
                                          &                            & 2021-2024 & \ding{55} & \ding{51} & \ding{51} & \ding{51} & \ding{51} & \ding{51} & \ding{51} & \ding{51} & \ding{51} & \ding{51} \cr 
    \cmidrule(lr){2-13}
                                          & \multirow{3}{*}{NI 225}    & 2019-2022 & \ding{55} & \ding{55} & \ding{55} & \ding{55} & \ding{55} & \ding{55} & \ding{55} & \ding{55} & \ding{55} & \ding{55} \cr 
                                          &                            & 2020-2023 & \ding{55} & \ding{55} & \ding{55} & \ding{55} & \ding{55} & \ding{55} & \ding{55} & \ding{55} & \ding{55} & \ding{55} \cr 
                                          &                            & 2021-2024 & \ding{55} & \ding{55} & \ding{55} & \ding{55} & \ding{55} & \ding{55} & \ding{55} & \ding{55} & \ding{55} & \ding{55} \cr 
    \bottomrule[1.5pt]
    \end{tabular}  
    \end{threeparttable}  
\end{table*}

\section{Conclusion}
\label{conclusion}

Due to the highly complex nonlinear structures often present in real-world data, linear Granger causality tests may perform poorly in such settings. In addition, causal relationships may exist at multiple lag orders. Although nonparametric causality tests are flexible, their applicability is limited by the curse of dimensionality. Therefore, it is meaningful to propose a testing method that can effectively handle complex nonlinear causality while also mitigating the effects of high dimensionality. This paper proposes a doubly robust test for Granger causality that integrates deep neural network estimators within an empirical process framework. The asymptotic properties of the test are established under the null hypothesis, alternative hypothesis, and a sequence of local alternatives. To support practical implementation, a multiplier bootstrap procedure is introduced, and its asymptotic validity is rigorously justified. The proposed test is specifically designed to handle the causality at multiple lags. A set of Monte Carlo simulations is conducted to evaluate the performance of the proposed test under finite samples. Finally, we apply the method to investigate potential nonlinear causality between stock prices and trading volume, yielding insightful empirical findings.

In our doubly robust Granger causality test, the types of deep neural networks used are relatively basic. More complex architectures may yield improved results, as in \cite{zhang2025doubly}. For instance, increasing the depth of the network or adopting more advanced neural network models could potentially enhance the test’s performance. Additionally, our parameter settings for the neural networks are relatively flexible. When applying the method to different causality testing problems, experienced practitioners might be able to fine-tune the model to obtain more insightful conclusions. In these respects, our work may open up potential directions for future research.

In the numerical simulations, the DGPs used in this paper still leave room for improvement. Due to the multi-lag structure of the processes, it is necessary to constrain the generated data using appropriate parameter settings; otherwise, divergence may occur, making the data unusable. This paper does not provide an in-depth study on the selection of these parameters. Therefore, one potential improvement is to conduct systematic testing and tuning of the parameters used in the DGPs. We believe that better parameter choices in the data generation process could more clearly demonstrate how increasing dimensionality affects the performance of Granger causality tests with multiple lags. This would not only benefit the method proposed in this paper, but also serve future developments in this line of research.

In addition, this paper focuses exclusively on testing Granger causality between two time series. We believe that the doubly robust Granger causality test exhibits sufficiently strong performance to be extended to multivariate time series settings—for example, testing whether both $X_t$ and $Z_t$ Granger-cause $Y_t$. Such an extension also opens the door to applications in richer data environments, such as panel data \cite{lu2017granger}. Naturally, however, increasing both the dimensionality of the time series and the number of lags will inevitably lead to greater computational burden and higher costs. Since the test relies on deep neural network estimators, its computational burden is greater than that of traditional nonparametric methods or tests involving only two time series. Therefore, improving the computational efficiency and scalability of the proposed test is another meaningful direction for future research. We leave these questions for further exploration.

\bibliographystyle{cas-model2-names}

\bibliography{cas-refs}

\appendix

\hypertarget{myappendix}{}
\section*{Appendix}
\label{appendixbegin}

This appendix first summarizes the assumptions in Appendix~\ref{assumptions}.
Appendices \ref{tech lemmas} and \ref{main proofs} then contain the technique lemmas, main proofs, respectively.

\section{Assumptions}
\label{assumptions}

\setcounter{equation}{0}
\renewcommand{\theequation}{A.\arabic{equation}}

\begin{assumption}
\label{betamixing}
Assume $X_t$ and $Y_t$ are stationary, and satisfy $\beta$-mixing condition, where the mixing coefficient $\beta(t) \leq C_1 \exp \left(-C_2 t\right)$ for some constants $C_1, C_2>0$. Suppose $\{Y_t - \mathbb{E}(Y_t|\boldsymbol{W}_{t-1}), \mathcal{F}_t\}$ is a martingale difference sequence.
\end{assumption}

\begin{assumption}
\label{boundv}
Suppose the following conditions hold. \\
(i) Let $\mathcal{X},\mathcal{Y}$ denote the support of $X_t,Y_t$. Suppose $\mathcal{X}$ and $\mathcal{Y}$ are compact sets of $\mathbb{R}$. \\
(ii) For an absolute constant $C_m>0$, assume $\left\|m\right\|_{\infty} \leq C_m$, and $\left\|f\right\|_{\infty} \leq C_m$ for any $f \in \mathcal{F}_{\mathrm{MLP}}$.
\end{assumption}

\begin{assumption}
\label{unboundv}
Suppose the density of $Y_t$ follows a sub-Gaussian distribution with parameter $\sigma_Y$. 
\end{assumption}

\begin{assumption}
\label{mlpf}
Assume that $m(\cdot)$ lies in the Sobolev ball with smoothness 
$$ \beta_0 \in \mathbb{N}_{+}: \left\{f: \max _{\boldsymbol{\beta},\|\boldsymbol{\beta}\|_1 \leq \beta_0} \operatorname*{ess\ sup}\limits_{x}\left|D^{\boldsymbol{\beta}} f(x)\right| \leq 1\right\},$$
where the maximum is taken over all $q$-dimensional non-negative integer-valued vectors $\boldsymbol{\beta}$ the sum of whose elements is no greater than $\beta_0$, and $D^{\boldsymbol{\beta}} f$ denotes the weak derivative (\citeauthor{gine2016mathematical}, \citeyear{gine2016mathematical}). 
\end{assumption}

\begin{assumption}
\label{smooth}
Suppose the following conditions hold for $f_{\boldsymbol{X}_{t-1} | \boldsymbol{Y}_{t-1}}(\cdot|\cdot)$.\\
(i) Suppose $f_{\boldsymbol{X}_{t-1} | \boldsymbol{Y}_{t-1}}(x | y)$ can be well-approximated by a conditional Gaussian mixture model with $G$ components, in that, there exists some constant $\omega_1>0$, such that
$$
\left\lvert f_{\boldsymbol{X}_{t-1} | \boldsymbol{Y}_{t-1}}(x | y)-\sum_{g=1}^G \frac{\alpha_g(y)}{\sqrt{2 \pi} \sigma_g(y)} \exp \left\{-\frac{\left[x-\mu_g(y)\right]^2}{2 \sigma_g^{2}(y)}\right\}\right\rvert =O\left(G^{-\omega_1}\right),
$$
where the big-O term is uniform in $x$ and $y$.\\
(ii) Suppose $\left\{\mu_g\right\}_{g=1}^G$ and $\left\{\sigma_g\right\}_{g=1}^G$ are uniformly bounded away from infinity, and there exist a constant $C>0, \omega_2 \geq 0$, such that $\sigma_g(y) \geq C G^{-\omega_2}$ for any $g$ and $y$.\\
(iii) Suppose $\alpha_g(\cdot), \mu_g(\cdot)$, and $\sigma_g(\cdot), g=1, \ldots, G$, all lie in the Sobolev ball with the smoothness 
$$\gamma_0 \in \mathbb{N}_{+}:\left\{f: \max_{\boldsymbol{\gamma},\| \boldsymbol{\gamma}\|_1 \leq \gamma_0} \sup_x \lvert D^{\boldsymbol{\gamma}} f(x) \rvert <+\infty\right\},$$
where the maximum is taken over all $q$-dimensional non-negative integer-valued vectors $\boldsymbol{\gamma}$ the sum of whose elements is no greater than $\gamma_0$, and $D^{\boldsymbol{\gamma}} f$ is the weak derivative.
\end{assumption}

\begin{assumption}
\label{mdncondition}
Suppose the following conditions hold for the MDN model.\\
(i) Suppose the MDN function class is given by, for some sufficiently large constant $C$,
$$
\begin{array}{r}
\mathcal{F}_{\mathrm{MDN}}=\Big\{f(x | y)=\sum\limits_{g=1}^G \frac{\alpha_g(y)}{\sqrt{2 \pi} \sigma_g(y)} \exp \left\{-\frac{\left(x-\mu_g(y)\right)^2}{2 \sigma_g^2(y)}\right\}: \inf _{x, y} f(x | y) \geq C^{-1}, \\
\sup _{y, g}\lvert \mu_g(y)\rvert  \leq C, C^{-1} G^{-\omega_2} \leq \inf _{y, g} \sigma_g(y) \leq \sup _{y, g} \sigma_g(y) \leq C\Big\},
\end{array}
$$
where $\alpha_g, \mu_g$ and $\sigma_g$ are parametrized via deep neural networks.\\
(ii) The total number of parameters $W$ in the MDN model is proportional to $G^{(q+\gamma_0) / \gamma_0} n^{q /(2 \gamma_0+q)}$ $\log (G n)$, where $\gamma_0$ is the smoothness parameter specified in Assumption \ref{smooth} (iii).
\end{assumption}

\begin{assumption}
\label{convergencerate}
Suppose the following conditions hold.\\
(i) Let $\widehat{m}$ denote the deep
MLP-ReLU network estimator estimator of $\mathbb{E}(Y_t|\boldsymbol{Y}_{t-1})$. Suppose $\widehat{m}$ converge at a rate of $O\left(T^{-\kappa_0}\right)$ for some $\kappa_0> 1/4$.
More specifically, suppose
$$
\mathbb{E}_n \left( \widehat{m} - m \right)^2 = O\left(n^{-2\kappa_0}\right),
$$
where $\mathbb{E}_n$ denotes the empirical expectation.\\
(ii) Suppose $\widehat{f}_{\boldsymbol{X}_{t-1}|\boldsymbol{Y}_{t-1}}$ converge at a rate of $O\left(T^{-\kappa_0}\right)$ for some $\kappa_0> 1/4$.
More specifically, suppose
$$
{\int_{x, y}\left\lvert\widehat{f}_{\boldsymbol{X}_{t-1}| \boldsymbol{Y}_{t-1}} (x | y)-f_{\boldsymbol{X}_{t-1} |\boldsymbol{Y}_{t-1}}(x | y)\right\rvert^2 \mathrm{d}x \; \mathrm{d}y} =O\left(n^{-2 \kappa_0}\right).
$$
\\
(iii) Suppose the number of samples generated by $\widehat{f}_{\boldsymbol{X}_{t-1}|\boldsymbol{Y}_{t-1}}$ satisfies $M=\kappa_1 n^{\kappa_2}$ for some $\kappa_1>0, \kappa_2 \geq 1 / 2$, and $q \leq \max \left(\rho_0 n, n-2\right)$ for some constant $0<\rho_0<1$.\\
(iv) Suppose the number of $(\mu, \nu)$ pairs $L$ grows polynomially fast with respect to $n$. 

\end{assumption}

\begin{assumption}
\label{distribution}
Suppose the following conditions hold.\\
(i) The density functions of $X_t$ and $Y_t$ exist and is bounded.\\
(ii) The density of $\varepsilon_t = Y_t - \mathbb{E}(Y_t | \boldsymbol{W}_{t-1})$ follows a sub-Gaussian distribution with parameter $\sigma_0$.
\end{assumption}

\begin{assumption}
\label{local_alternative}
Suppose the following conditions hold.\\
(i) $\Delta(\boldsymbol{w})$ is uniformly bounded with respect to $\boldsymbol{w}$.\\
(ii) $(n-q)^{-1} \sum\limits_{t=q+1}^n  \Delta\left(\boldsymbol{W}_{t-1}\right) \xrightarrow{\text { a.s. }} Q(\mu, \nu) =\mathbb{E}\left[ \Delta\left(\boldsymbol{W}_{t-1}\right) \right]$ uniformly in $(\mu, \nu) \in \mathcal{W}$.
\end{assumption}

\section{Proofs of lemmas}
\label{tech lemmas}

\setcounter{equation}{0}
\renewcommand{\theequation}{B.\arabic{equation}}

\setcounter{lemma}{0}

\subsection{Lemma \ref{mlpL2}}
\begin{lemma}
Suppose Assumptions \ref{betamixing}, \ref{boundv} and \ref{mlpf} hold. Then, there exists a MLP estimator $\widehat{m}(\boldsymbol{Y}_{t-1})$ for $m(\boldsymbol{Y}_{t-1})$, with width $H_n \asymp n^{\frac{q}{2(\beta_0+q)}} \log ^2 (n)$ and depth $L_n \asymp \log (n)$, such that
$$
\mathbb{E} \left( \widehat{m}(\boldsymbol{Y}_{t-1})-m(\boldsymbol{Y}_{t-1})\right)^2 \leq C \cdot\left\{n^{-\frac{\beta_0}{\beta_0+q}} \log ^9 (n)+\frac{\log (n) \log \log (n)}{n}\right\},
$$
and
$$
\mathbb{E}_n \left( \widehat{m}(\boldsymbol{Y}_{t-1})-m(\boldsymbol{Y}_{t-1})\right)^2 \leq C \cdot\left\{n^{-\frac{\beta_0}{\beta_0+q}} \log ^9 (n)+\frac{\log (n) \log \log (n)}{n}\right\},
$$
for a constant $C>0$, with probability at least $1-4\exp \left(-n^{\frac{q}{\beta_0+q}} \log ^8 (n)\right)-O(n^{-1})$, where $\mathbb{E}_n$ denotes the empirical expectation.
\end{lemma}

\subsection{Lemma \ref{mlpL2un}}
\begin{lemma}
Suppose Assumptions \ref{betamixing},  \ref{unboundv} and \ref{mlpf} hold. Suppose the MLP is specified with width
$H_n \asymp n^{\left(\frac{q}{\beta_0+q}\right)\left(1 / 2-\kappa_B\right)} \log ^2(n)$ and depth $L_n \asymp \log (n)$. Let $
K_n=n^{-\left(\frac{\beta_0}{\beta_0+q}\right)\left(1 / 2-\kappa_B\right)} \log ^6(n)$.
Then, for the MLP estimator $\widehat{m}(\boldsymbol{Y}_{t-1})$, 
there exist constants $C, C_1, C_2>0$ independent of $n$, for all $n$ sufficiently large, such that
$$
\begin{aligned}
& P\left(\mathbb{E}\left|\widehat{m}(\boldsymbol{Y}_{t-1})-m(\boldsymbol{Y}_{t-1})\right|^2 \leq C^2 K_n^2\right) \geq 1-e^{-n\left(\frac{\beta_0}{\beta_0+q}\right)^{\left(1 / 2-\kappa_B\right)}}-\frac{2 C_1 n^{1-C_2 \log (n)}}{\log (n)}-o(1/n), \\
& P\left(\mathbb{E}_n\left |\widehat{m}(\boldsymbol{Y}_{t-1})-m(\boldsymbol{Y}_{t-1})\right |^2 \leq C^2 K_n^2\right) \geq 1-4 e^{-n\left(\frac{\beta_0}{\beta_0+q}\right)^{\left(1 / 2-\kappa_B\right)}}-\frac{12 C_1 n^{1-C_2 \log (n)}}{\log (n)}-o(1/n).
\end{aligned}
$$

\end{lemma}

\subsection{Lemma \ref{mdnL2}}
\begin{lemma}
    Suppose Assumptions \ref{betamixing} and \ref{smooth} hold.  Then, there exists a certain MDN function class satisfying Assumption \ref{mdncondition}, such that the resulting MDN estimator $\widehat{f}_{\boldsymbol{X}_{t-1} | \boldsymbol{Y}_{t-1}}(x|y)$ satisfies that
$$
\begin{aligned}
\mathbb{E}\left( \widehat{f}_{\boldsymbol{X}_{t-1} | \boldsymbol{Y}_{t-1}}(x|y)-f_{\boldsymbol{X}_{t-1} | \boldsymbol{Y}_{t-1}}(x|y)\right)^2 & ={\int_{x, y}\lvert \widehat{f}_{\boldsymbol{X}_{t-1} | \boldsymbol{Y}_{t-1}}(x | y)-f_{\boldsymbol{X}_{t-1} | \boldsymbol{Y}_{t-1}}(x | y)\rvert ^2 d x d y} \\
& \leq C q\left\{G^{-\omega_1}+G^{\frac{\gamma_0+q}{2 \gamma_0}+4 \omega_2} n^{-\frac{\gamma_0}{2 \gamma_0+q}} \log ^3(n G)\right\},
\end{aligned}
$$
for some constants $C, \omega_1, \omega_2>0$, with probability at least $1-O\left(n^{-1}\right)$.
\end{lemma}

\subsection{Lemma \ref{phiL2}}

\begin{lemma}
    Suppose Assumptions \ref{betamixing} and \ref{convergencerate} (ii)-(iv) hold. Then, there exists a constant $\kappa>1 / 2$, such that
$$
\begin{gathered}
\max _{1 \leq l \leq L} \int_y\left\lvert\widehat{\phi}\left(\nu_l | y\right)-\phi\left(\nu_l | y\right)\right\rvert^2 \mathbb{F}(\mathrm{d} y)=O_p\left(n^{-\kappa}\right) \\
\end{gathered}
$$
where $\mathbb{F}$ denotes the cumulative distribution function of $\boldsymbol{Y}_{t-1}$, and ${\phi}(\nu|y) = \mathbb{E}\left(e^{i\nu^\top \boldsymbol{X}_{t-1}}|\boldsymbol{Y}_{t-1}=\boldsymbol{y}\right)$.
\end{lemma}

\section{Main proofs}
\label{main proofs}

\setcounter{equation}{0}
\renewcommand{\theequation}{C.\arabic{equation}}

\setcounter{proposition}{0}

\subsection{Proof for Proposition \ref{hyp=}}
\begin{proposition}
\label{hyp=}
The following testing hypotheses are equivalent:
\begin{align*}
& \mathrm{H_0}: \mathbb{E}\left[\left(Y_t - m(\boldsymbol{Y}_{t-1})\right)e^{i{w}^\top\boldsymbol{W}_{t-1}}\right] = 0,\ \  \forall {w}\in \mathcal{W}, \mathrm{a.s.} \\
& \mathrm{H_1}: \mathbb{E}\left[\left(Y_t - m(\boldsymbol{Y}_{t-1})\right)e^{i{w}^\top\boldsymbol{W}_{t-1}}\right] \neq 0,\ \  \forall {w}\in \mathcal{W}, \mathrm{a.s.}
\end{align*}
and
\begin{align*}
& \mathrm{H_0}: \mathbb{E} \left[ \Big(Y_t-{m}(\boldsymbol{Y}_{t-1})\Big)e^{i\mu^\top\boldsymbol{Y}_{t-1}}\left(e^{i\nu^\top\boldsymbol{X}_{t-1}}-{\phi}(\nu| \boldsymbol{Y}_{t-1})\right) \right]=0, \mathrm{a.s.} \\
& \mathrm{H_1}: \mathbb{E} \left[ \Big(Y_t-{m}(\boldsymbol{Y}_{t-1})\Big)e^{i\mu^\top\boldsymbol{Y}_{t-1}}\left(e^{i\nu^\top\boldsymbol{X}_{t-1}}-{\phi}(\nu| \boldsymbol{Y}_{t-1})\right) \right]\neq 0, \mathrm{a.s.}
\end{align*}
\end{proposition}

\subsection{Proof for Proposition \ref{mainpro}}

\begin{proposition}
\label{mainpro}
Suppose Assumptions \ref{betamixing}, \ref{convergencerate} and \ref{distribution} hold. Then, under the null hypothesis,
$$
\sup_{(\mu, \nu) \in \mathcal{W}} \left\lvert S_n(\mu, \nu)-\widehat{S}_n(\mu, \nu)\right\rvert =o_p(1),
$$
where ${S}_n({\mu},{\nu})$ and $\widehat{S}_n({\mu},{\nu})$ are defined in \eqref{infedr} and \eqref{fedr}, respectively.
\end{proposition}

\setcounter{theorem}{0}

\subsection{Proof for Theorem \ref{theoremnull}}
\begin{theorem}
Suppose Assumptions \ref{betamixing}, \ref{convergencerate} and \ref{distribution} hold. Then under the null,
$$
\widehat{S}_n(\mu, \nu) \rightsquigarrow S_{\infty}(\mu, \nu),
$$
where $S_{\infty}(\cdot, \cdot)$ is a zero mean Gaussian process with covariance kernel $\mathbb{E}\left[S_{\infty}(\mu, \nu) S_{\infty}\left({\mu}', {\nu}'\right)\right]$.
\end{theorem}

\subsection{Proof for Corollary \ref{conull}}
\setcounter{corollary}{0}
\begin{corollary}
Suppose Assumptions \ref{betamixing}, \ref{convergencerate} and \ref{distribution} hold. Then under the null,
$$
\mathrm{KS}_n \rightsquigarrow \mathrm{KS}_{\infty}=\sup _{(\mu,\nu) \in \mathcal{W}} \max\left(| S_{R}(\mu, \nu)| , | S_{I}(\mu, \nu)| \right),
$$
where $S_{R}(\mu, \nu)$ and $S_{I}(\mu, \nu)$ denote the real and imaginary part of $S_{\infty}$ defined in Theorem 1.
\end{corollary}

\subsection{Proof for Theorem \ref{theoremal}}

\begin{theorem}
Suppose Assumptions \ref{betamixing}, \ref{convergencerate} and \ref{distribution} hold. Then under the alternative, for each $(\mu, \nu) \in \mathcal{W}$,
$$
n^{-1/2} \widehat{S}_n(\mu, \nu) \stackrel{P}{\rightarrow} \mathbb{E}\left[\left(Y_t-m(\boldsymbol{Y}_{t-1})\right)e^{iw^\top \boldsymbol{W}_{t-1}}\right].
$$
\end{theorem}

\subsection{Proof for Theorem \ref{theoremlo}}

\begin{theorem}
Suppose Assumptions \ref{betamixing}, \ref{convergencerate}, \ref{distribution} and \ref{local_alternative} hold. Then under the sequence of local alternatives,
$$
\widehat{S}_n(\cdot, \cdot) \rightsquigarrow S_{\infty}(\cdot, \cdot)+Q(\cdot, \cdot),
$$
where $S_{\infty}(\mu, \nu)$ is the Gaussian process defined in Theorem \ref{theoremnull} and $Q(\cdot, \cdot)$ is the shift function.
\end{theorem}

\subsection{Proof for Corollary \ref{colo}}
\begin{corollary}
Suppose Assumptions \ref{betamixing}, \ref{convergencerate}, \ref{distribution} and \ref{local_alternative} hold. Then, under the sequence of local alternatives,
$$
\begin{gathered}
\mathrm{KS}_n \rightsquigarrow \sup _{(\mu, \nu) \in \mathcal{W}}\lvert S_{\infty}(\mu, \nu)+G(\mu, \nu)\rvert ,
\end{gathered}
$$
where $S_{\infty}(\mu, \nu)$ and $G(\mu, \nu)$ are defined in Theorem \ref{theoremlo}.
\end{corollary}

\subsection{Proof for Proposition \ref{propboo}}
\begin{proposition}
\label{propboo}
Suppose Assumptions \ref{betamixing}, \ref{convergencerate} and \ref{distribution} hold. The multiplier bootstrap empirical process satisfies
$$
\sup _{(\mu, \nu) \in \mathcal{W}}\lvert S_n^*(\mu, \nu)-\widehat{S}_n^*(\mu, \nu)\rvert =o_p(1),
$$
where
$$
\begin{aligned}
    \widehat{S}_n^*({\mu},{\nu}) =& \frac{1}{\sqrt{n-q}} \mathop{\sum}\limits_{t=q+1}^n \xi_t\left(Y_t-\widehat{m}(\boldsymbol{Y}_{t-1})\right)e^{i\mu^\top\boldsymbol{Y}_{t-1}}\left(e^{i\nu^\top\boldsymbol{X}_{t-1}}-\widehat{\phi}(\nu| \boldsymbol{Y}_{t-1})\right), \\
    {S}_n^*({\mu},{\nu})=& \frac{1}{\sqrt{n-q}} \mathop{\sum}\limits_{t=q+1}^n \xi_t\Big(Y_t-{m}(\boldsymbol{Y}_{t-1})\Big)e^{i\mu^\top\boldsymbol{Y}_{t-1}}\left(e^{i\nu^\top\boldsymbol{X}_{t-1}}-{\phi}(\nu| \boldsymbol{Y}_{t-1})\right),
\end{aligned}
$$
and $\{\xi_t\}$ is i.i.d standard normal random variables.
\end{proposition}

\subsection{Proof for Theorem \ref{theoremboo}}
\begin{theorem}
Suppose Assumptions \ref{betamixing}, \ref{convergencerate} and \ref{distribution} hold. Then, under the null, under the alternative, or under the sequence of local alternatives,
$$
\widehat{S}_n^*(\mu, \nu) \underset{*}{\stackrel{P}{\rightarrow}} S_{\infty}(\mu, \nu)
$$
where $S_{\infty}(\mu, \nu)$ is the Gaussian process defined in Theorem \ref{theoremnull}, and $\underset{*}{\stackrel{P}{\rightarrow}}$ denotes weak convergence in probability under the bootstrap law, conditional on $\left\{\left(X_t^{\top}, Y_t^{\top}\right)^{\top}\right\}_{t=1}^n$. In addition, $\mathrm{KS}_n^* \underset{*}{\rightsquigarrow} \mathrm{KS}_{\infty}$ with $\mathrm{KS}_{\infty}$ defined in Corollary \ref{conull}.
\end{theorem}

\end{document}